\documentclass[12pt]{article}
\usepackage{amsmath,amsfonts,amssymb,mathtools}
\usepackage[nosort]{cite}
\usepackage{hyperref}
\usepackage{color}
\usepackage{epsfig}
\usepackage{psfrag}	
\usepackage{tikz}
\usepackage{slashed}
\usepackage{tensor}
\usepackage{ulem}
\usepackage[font=small,labelfont=bf,width=1\textwidth]{caption}
\usepackage{comment}
\usepackage[all]{xy}
\usepackage{multirow}
\usepackage{float}
\usepackage{cite}
\usepackage{color}

\usepackage{tocloft} % Layout of table of contents
\usepackage{pifont}
\newcommand{\cmark}{\ding{51}}
\newcommand{\xmark}{\ding{55}}
\usepackage{xltabular}
\usepackage{ulem}
\usepackage{soul,xcolor}
\setstcolor{red}

\numberwithin{equation}{section}

\usepackage{bbold}
\DeclareFontFamily{U}{futm}{}
\DeclareFontShape{U}{futm}{m}{n}{
  <-> s * [.97534] fourier-bb % but changing the magnification factor
  }{}
\DeclareMathAlphabet{\mathbbb}{U}{futm}{m}{n}

\usepackage{graphics}
\newcommand{\Db}{\reflectbox{$\mathbb{D}$}}
\newcommand{\Lambdab}{\text{\reflectbox{$\mathbb{\Lambda}$}}}
\newcommand{\Sigmab}{\text{\reflectbox{$\mathbb{\Sigma}$}}}
\newcommand{\Ob}{\text{\reflectbox{$\mathbb{O}$}}}
\newcommand{\Fb}{\text{\reflectbox{$\mathbb{F}$}}}
\newcommand{\bOb}{{\mathbbb{O}}}
\newcommand{\bDb}{{\mathbbb{D}}}
\newcommand{\bLambdab}{\text{$\mathrlap{\mathbb{\Lambda}}\reflectbox{$\mathbb{\Lambda}$}$}}
\newcommand{\llangle}{\big\langle\hspace{-1.2mm}\big\langle\hspace{-.5mm}}
\newcommand{\rrangle}{\hspace{-.5mm}\big\rangle\hspace{-1.2mm}\big\rangle}
\newcommand{\LLangle}{\left\langle\hspace{-2mm}\left\langle\hspace{-.5mm}}
\newcommand{\RRangle}{\hspace{-.5mm}\right\rangle\hspace{-2mm}\right\rangle}
\DeclareMathOperator{\Tr}{Tr}

\DeclareMathOperator{\diag}{diag}

\def\bF{{\mathbb{F}}}

\def\bC {\mathbb{C}}

\def\bO {\mathbb{O}}
\def\bP {\mathbb{P}}
\def\bLambda {\mathbb{\Lambda}}
\def\bSigma{\mathbb{\Sigma}}

\def\bR {\mathbb{R}}

\newcommand{\bea}{\begin{eqnarray}}
\newcommand{\eea}{\end{eqnarray}}
\newcommand{\beq}{\begin{equation}}
\newcommand{\eeq}{\end{equation}}
\newcommand{\bal}{\begin{equation}\begin{aligned}}
\newcommand{\eal}{\end{aligned} \end{equation}}

\newcommand{\address}[1]{\vbox{\center\em#1}}
\renewcommand{\title}[1]{\vbox{\center\huge{#1}}\vspace{5mm}}

\newcommand{\bD}{{\mathbb D}}
\newcommand{\cA}{{\mathcal A}}

\newcommand{\cD}{{\mathcal D}}

\newcommand{\cG}{{\mathcal G}}

\newcommand{\cL}{{\mathcal L}}
\newcommand{\cM}{{\mathcal M}}
\newcommand{\cN}{{\mathcal N}}
\newcommand{\cP}{{\mathcal P}}

\newcommand{\cO}{{\mathcal O}}

\newcommand{\su}{\mathfrak{s}\mathfrak{u}}
\newcommand{\sof}{\mathfrak{s}\mathfrak{o}}

\newcommand{\osp}{{\mathfrak{osp}}}
\newcommand{\unitary}{{\mathfrak{u}(1)}}

\begin{document}

\begin{titlepage}
\begin{center}

\vspace*{20mm}

\title{1/3 BPS loops and defect CFTs in ABJM theory}

\vspace{7mm}

\renewcommand{\thefootnote}{$\alph{footnote}$}

Nadav Drukker%
\footnote{\href{mailto:nadav.drukker@gmail.com}
{\tt nadav.drukker@gmail.com}}
and Ziwen Kong%
\footnote{\href{mailto:ziwen.kong@kcl.ac.uk}
{\tt ziwen.kong@kcl.ac.uk}}

\vskip 2mm
\address{
Department of Mathematics, King's College London,
\\
The Strand, WC2R 2LS London, United-Kingdom}

\renewcommand{\thefootnote}{\arabic{footnote}}
\setcounter{footnote}{0}

\end{center}
\vspace{4mm}
\abstract{
\normalsize{
\noindent
We address a longstanding question of whether ABJM theory has 
Wilson loop operators preserving eight supercharges (so 1/3 BPS). 
We present such Wilson loops made of a large supermatrix combining 
two 1/2 BPS Wilson loops. We study the spectrum of operator insertions 
into them including the displacement operator and several others 
and study their correlation functions. Another natural construction 
arising in this context are Wilson loops with 
alternating superconnections. This amounts to including 
``defect changing operators'' along the loop, 
similar to a discrete cusp. This insertion is topological 
and preserves two supercharges. We study the multiplet of this 
operator and how it can be used to introduce further operators. 
We also construct the defect conformal manifold arising from 
marginal defect operators.
}}
\vfill

\end{titlepage}

\section{Introduction}
\label{sec:intro}

ABJM theory~\cite{Aharony:2008ug} has a rich spectrum of line operators including 
the 1/2 BPS loop~\cite{Drukker:2009hy} and 1/6 BPS loops. The latter may be 
bosonic~\cite{Drukker:2008zx,Chen:2008bp,Rey:2008bh} with only a single gauge 
field or include fermi fields like the 1/2 BPS 
ones~\cite{Ouyang:2015iza,Ouyang:2015bmy, Mauri:2018fsf, Drukker:2019bev}. 
There are also Wilson loops preserving fewer 
supercharges~\cite{Cardinali:2012ru, drukker:2020bps, Drukker:2020dvr, 
Drukker:2022ywj, Drukker:2022bff, Castiglioni:2022yes}, though they 
are not conformal. Finally there are vortex loops~\cite{Drukker:2008jm} that are 
1/2 BPS or 1/3 BPS (though there should also be less supersymmetric versions).

A natural question that many experts have tried to address, is whether there are 
also 1/3 BPS Wilson loops in this theory. Given that a vortex loop exists, there 
is an appropriate superalgebra. Indeed, the $\osp(6|4)$ superalgebra 
of ABJM is broken to $\su(1,1|3)%\oplus\unitary
$ by the 1/2 BPS line and to 
$\su(1,1|1)\oplus\su(2)\oplus\unitary$ by the 1/6 BPS loops (for the bosonic loop, 
the $\unitary$ is enhanced to another $\su(2)$). The 1/3 BPS algebra is 
$\su(1,1|2)\oplus\unitary$. This latter algebra (up to a $\unitary$ factor) is 
also the symmetry of the 1/2 BPS loops of $\cN=4$ Chern-Simons theories, 
which is a hint for our construction.

The 1/2 BPS Wilson loops of $\cN=4$ theories~\cite{Ouyang:2015qma, 
Cooke:2015ila} have a degeneracy of pairs of loops preserving the same 
eight supercharges. Choosing then eight of the twelve supercharges of a 
1/2 BPS loop $W_1^+$ of ABJM that generate an $\su(1,1|2)$ subalgebra, there 
should be another Wilson loop, $W_4^-$ preserving the same supercharges. This 
second Wilson loop is also 1/2 BPS, but the linear combination 
$W_{1/3}=n_1W_1^++n_4W_4^-$ 
is 1/3 BPS. Explicit expressions for $W_1^+$ and $W_4^-$ are presented in 
Appendix~\ref{app:WLs}: \eqref{L1}, \eqref{L4}.

Defining an operator as a linear combination of other ones may not seem 
fundamental, and one may raise the objection that they should each 
be studied independently. One way to see that this is not the case 
is that this linear combination arises naturally when considering 
Wilson loops based on superconnections larger than that of the 
1/2 BPS loop. Such larger constructions with repeated entries from 
the same gauge field were proposed 
in~\cite{drukker:2020bps, Drukker:2020dvr, Drukker:2022ywj} and give 
rise also to operators that cannot be expressed in the block-diagonal 
form of $n_1W_1^++n_4W_4^-$. While the 1/3 BPS loop itself can 
be written this way, it can be deformed into non-diagonal loops, 
so operator insertions into $W_{1/3}$ cannot be factorised as 
insertions into $W_1^+$ and those in $W_4^-$.

Having realised this 1/3 BPS line, we turn to studying its properties 
and in particular the defect CFT for operator insertions along 
it~\cite{Drukker:2006xg, Billo:2016cpy, Cooke:2017qgm, Giombi:2017cqn, 
Bianchi:2020hsz}. Part of this analysis relies on the explicit 
realisation of $W_{1/3}$ presented here and part is based 
on representation theory of the superconformal group, so 
is valid for any 1/3 BPS line operator including the vortex loop 
of~\cite{Drukker:2008jm} or any further line operators that may 
be found in the future.

The displacement and tilt operators are insertions 
that arise from broken translation and R-symmetry, respectively. 
As reviewed in Appendix~\ref{app:CBD}, the two point functions 
of these operators are related to bremsstrahlung 
functions~\cite{Correa:2012at,Fiol:2012sg}. 
ABJM theory has a rich spectrum of such 
functions~\cite{Griguolo:2012iq, Bianchi:2017afp, 
Bianchi:2017ujp, Bianchi:2018scb}, and the case of the 1/3 
BPS Wilson loop is even richer.

Any 1/3 BPS line breaks the conformal group 
$\sof(4,1)\to\sof(2,1)\oplus\sof(2)$, just as the 1/2 BPS 
or any other conformal line operator, so has two displacement 
operators from the broken translations. The $\su(4)$ R-symmetry 
is broken to $\su(2)\oplus\unitary^2$ with five different pairs 
of tilt operators: a conjugate pair denoted $\bOb$ and 
$\bar\bOb$ for the broken generators between the two $\unitary$s 
and the others relating each of them to $\su(2)$ (they are denoted as 
$\bO^a$, $\bar\bO_a$, $\Ob_a$, $\bar\Ob^a$ with $a\in\{2,3\}$). 
These are really different operators 
with different normalisations, i.e. there are three 
bremsstrahlung functions related to R-symmetry 
breaking in addition to the one for a real cusp.

Following~\cite{Bianchi:2018scb,Bianchi:2018zpb}, we use in 
Section~\ref{sec:2pt} Ward identities and the explict form of the tilt operators 
to find relations between the bremsstrahlung functions. 
$\bOb$ is in the same multiplet with the displacement $\bDb$, so 
the associated bremsstrahlung functions are clearly related. For the other 
tilt operators, their sum is equal to that of $\bOb$.

In the case when the bremsstrahlung functions for $\bO^a$ and $\Ob_a$ are 
equal (for $n_1=n_4$), 
they are half of that of $\bOb$ or $\bDb$. A similar relation exists between 
the bremsstrahlung functions for the tilt and displacement of the bosonic loop, 
but here we find a very simple setting of the same phenomenon and a far 
easier proof of it.

Another natural object that arises in this context is a permutation 
operator, which we denote by $\sigma$, that replaces the connection 
of $W_1^+$ with that of $W_4^-$. This is the most clear manifestation of the 
nontrivial interplay between the two Wilson loops. This operator 
preserves two supercharges and has vanishing conformal dimension, 
so is topological. This enriches the spectrum of protected operators 
on $W_{1/3}$, but can also be considered as an operator on the 1/2 BPS 
$W_1^+$. We study some of its properties, but leave most of them 
for future work.

Going back to the five pairs of tilt operators, they are exactly 
marginal operators on $W_{1/3}$ and in Section~\ref{sec:DCM} we 
study the deformation of the loop by them. We 
follow~\cite{Drukker:2022pxk} to calculate the geometry of the 
resulting defect conformal manifold and precisely match it with 
the quotient $SU(4)/S(U(2)\times U(1)\times U(1))$.

Some background information and many technical details are relegated 
to the appendices.

\section{Realising 1/3 BPS Wilson loops}
\label{sec:gauging}

We construct here 1/3 BPS Wilson loops in ABJM theory and then recall 
some subtle features of general loops in this theory that play an 
important role for these new loops.

The simplest 1/2 BPS loop $W_1^+$ is formed out of $\cL_1^+$ \eqref{L1}
\beq
\label{W1}
W_{1}^+=\Tr\cP\exp\int_{-\infty}^\infty 
i\cL_1^+\,dx\,.
\eeq
For $SU(N_1)\times SU(N_2)$ ABJ(M) theory, the superconnection 
$\cL_1^+$ is an $SU(N_1|N_2)$ matrix. Here we take it to be a straight 
line in the $x^3$ direction (which we denote as $x$). 
For the circular loop there is subtlety of taking the trace or 
supertrace~\cite{Drukker:2009hy, Drukker:2019bev}, but here we are 
taking the infinite line, so really it is an open line. We write trace, 
and if adapting to a circle, one should include a twist operator 
\eqref{T} or following the convensions of~\cite{Drukker:2019bev}, 
use a supertrace.

Another 1/2 BPS Wilson line is $W_4^-$, made out of the 
superconnection $\cL_4^-$ \eqref{L4}. Each of $W_1^+$ and 
$W_4^-$ preserves twelve supercharges, and when they are 
along the same line, they have eight supercharges in common.

To define the 1/3 BPS loop, we take a bigger structure combining both 
superconnections
\beq
\label{trivial1/3}
W_{1/3}=\Tr\cP\exp i\int_{-\infty}^\infty 
\diag(\underbrace{\cL_1^+,\cdots,\cL_1^+}_{n_1},
\underbrace{\cL_4^-,\cdots,\cL_4^-}_{n_4})\,dx\,.
\eeq
To avoid confusion, each $\cL_i^\pm$ is an $(N_1|N_2)$ supermatrix, 
so this is not a $(2n_1 N_1|2n_4 N_2)$ supermatrix, but 
rather $((n_1+n_4)N_1|(n_1+n_4)N_2)$. With this 
diagonal structure, this loop can also be written as 
$n_1 W_1^++n_4 W_4^-$.

This loop on its own is 1/3 BPS, resolving this long-standing question. 
One may wonder whether there are other 1/3 BPS loops with non-diagonal 
structure, as there are many constructions of BPS Wilson loops that do 
not respect it. We made an extensive and systematic search, based on the 
techniques of~\cite{Drukker:2020dvr, Drukker:2022ywj} and 
all the 1/3 BPS loops we found could be diagonalised to 
\eqref{trivial1/3}.

This construction of Wilson loops out of supermatrices has an 
$S(GL(n_1+n_4)\times GL(n_1+n_4))$ global symmetry, 
which was pointed out in~\cite{Drukker:2019bev, drukker:2020bps}. 
If we reorder the superconnection in a way that all $A^{(1)}_x$ are 
at the top left and $A^{(2)}_x$ at the bottom right, this group 
acts by independently rotating the $n_1+n_4$ copies of $A^{(1)}_x$ 
and of $A^{(2)}_x$, see \cite{drukker:2020bps} for details.
This global symmetry is important in the analysis of the space of 
BPS Wilson loops, as the Wilson loop is a trace, so 
we should really identify operators related by conjugation. 
It is also this action that 
allows us to diagonalise all other 1/3 BPS loops we found to the same 
form as \eqref{trivial1/3}.

In general, this action is not a local symmetry. The simplest manifestation 
of that is in the case of a single $\cL_1^+$, where the group is 
simply $GL(1)=\bC^*$ and it acts on the $2\times2$ structure within 
$\cL_1^+$ \eqref{L1} as conjugation by elements like
\beq
\label{T}
T=\begin{pmatrix}
I_{N_1}&0\\0&-I_{N_2}
\end{pmatrix}.
\eeq
This has the effect of changing the signs $\alpha\to-\alpha$, 
$\bar\alpha\to-\bar\alpha$. The local action of this operator was studied 
in~\cite{Gorini:2022jws}, where it was found to be a nontrivial 
operator in the defect CFT of the 1/2 BPS line. Though it has 
vanishing classical dimension, it is not BPS, so its dimension 
receives quantum corrections.

We focus instead on a diagonal $SL(n_1+n_4)$ subgroup 
which acts simultaneously on $A^{(1)}_x$ 
and $A^{(2)}_x$, not modifying $\cL_1^+$ and $\cL_4^-$. This is 
the obvious group acting by conjugation on the 
$n_1+n_4$ matrix of superconnections in \eqref{trivial1/3}.

\subsection{Permutation operators}
\label{sec:permutation}

Of the diagonal $SL(n_1+n_4)$ action on the matrix in 
\eqref{trivial1/3}, an $S(GL(n_1)\times GL(n_4))$ 
subgroups is in fact a local symmetry, as the 
superconnection is proportional to the identity 
in those blocks. Other group elements change the form 
of the connection and are nontrivial operations on 
the Wilson line and can be viewed as operators in a 
1d defect CFT.

To keep the gauge fields on the diagonal, the group elements we 
employ are permutations, changing the order of the entries. 
Explicitly for the case of $n_1=n_4=1$, there is a single non-trivial 
permutation
\beq
\label{sigma}
\sigma=\begin{pmatrix}
0&I_{N_1+N_2}\\I_{N_1+N_2}&0
\end{pmatrix},
\qquad
\sigma
\begin{pmatrix}
\cL_1^+&0\\0&\cL_4^-
\end{pmatrix}
\sigma
=\begin{pmatrix}
\cL_4^-&0\\0&\cL_1^+.
\end{pmatrix}
\eeq
As both $\cL_1^+$ and $\cL_4^-$ have the same gauge-group 
structure, there is no obstruction to doing this, and it 
is particularly nice since $W_1^+$ and $W_4^-$ share eight 
supercharges. Furthermore, as we show in 
Appendix~\ref{app:sigma-mult}, this combination preserves 
half the supercharges shared by the two lines. 

Another natural operator is $\tau=\diag(I,-I)$, 
satisfying $\tau\sigma\tau=-\sigma$. This is different from $T$ of 
\eqref{T}, which acts within a single block of these matrices, so on 
a single $\cL_i^{\pm}$. In this setting there are then two basic 
$T$-like operators, $\diag(T,1)$, $\diag(1,T)=\sigma\diag(T,1)\sigma$, 
and one can also multiply them with $\sigma$ and $\tau$.

Unlike $T$, the permutation $\sigma$ and $\tau\sigma$ are protected local 
operators. As their conformal dimension vanishes, they are topological and 
correlation functions of any other operators do not depend on the 
exact position where the permutation happens, as long as 
it does not cross any of the other operators.

$\sigma$, $T$ and $\tau$ are ``line changing operators'', 
similar to boundary changing operators in 2d CFTs.%
\footnote{It is also natural to relate $\sigma$ to permutation branes, 
see e.g.~\cite{Recknagel:2002qq}.}
The most studied 
such operators in Wilson lines are cusps, where the direction of the 
line changes, or there is a change in some internal 
parameters~\cite{Polyakov:1980ca, Korchemsky:1987wg, Drukker:1999zq, 
Drukker:2011za,Griguolo:2012iq}. 
Indeed in both $\cN=4$ SYM in 4d and in ABJM theory those were studied 
extensively and some cusps were shown to be 
BPS~\cite{Zarembo:2002an, Drukker:2011za, Griguolo:2012iq}. 
$T$ is similar to a non-BPS 
cusp and $\sigma$ to the BPS cusp. But unlike the usual cusps, 
both $T$ and $\sigma$ are discrete operations, so one cannot study 
them in a small or large angle expansion.

\subsection{1/2 BPS loop with alternating superconnections}
\label{sec:1/2sigma}
The operation of replacing part of a line with another connection arises 
naturally from the permutation symmetry above, but does not require large 
supermatrices. To see that consider
\beq
\label{sigmapm}
\sigma^+=\frac{1}{2}(1+\tau)\sigma=\begin{pmatrix}0&I\\0&0\end{pmatrix},
\qquad
\sigma^-=\frac{1}{2}(1-\tau)\sigma=\begin{pmatrix}0&0\\I&0\end{pmatrix}.
\eeq
Clearly $(\sigma^+)^2=(\sigma^-)^2=0$, so we should avoid that. On the 
other hand, $\sigma^+\sigma^-$ and $\sigma^-\sigma^+$ are projectors 
on the top or bottom parts of $\diag(\cL_1^+,\cL_4^-)$. Inserting this 
into the 1/3 BPS line we find
\beq
\label{1/3to1/2}
W_{1/3}[\sigma^+\sigma^-(0)]
=\Tr\cP e^{i\int_{-\infty}^0
\Big(\begin{smallmatrix}\cL_1^+(x)&0\\0&\cL_4^-(x)\end{smallmatrix}\Big)dx}
\begin{pmatrix}I&0\\0&0\end{pmatrix}
\cP e^{i\int_{0}^\infty
\Big(\begin{smallmatrix}\cL_1^+(x)&0\\0&\cL_4^-(x)\end{smallmatrix}\Big)dx}
=W_1^+\,.
\eeq
So this reduces the 1/3 BPS loop to the 1/2 BPS one. Inserting 
$\sigma^-\sigma^+$ reproduces $W_4^-$.

As stated, both $\sigma$ and $\tau$ are protected topological 
operators, and hence also $\sigma^\pm$. We can therefore 
separate the two insertions and in particular move $\sigma^-$ 
to $x\to\infty$, leaving us with the operator
\beq
\label{Wsigma}
W_{1/3}[\sigma^+(0)]=\Tr\cP\left[
\exp\int_{-\infty}^0 i\cL_1^+(x)\,dx
\,
\exp\int_0^\infty i\cL_4^-(x)\,dx
\right]
=W_1^+[\sigma].
\eeq
This loop starts with a single superconnection $\cL_1^+$ and switches 
at $x=0$ to $\cL_4^-$. In the last expression, we employed the notation 
$W_1^+[\sigma]$, where $\sigma$ is now an insertion in the 1/2 BPS loop 
that changes the connection. We can then denote $\sigma^-\simeq\bar\sigma$, 
where it is assumed that there is first a $\sigma$ insertion. On it's 
own $\bar\sigma$ is a good insertion in $W_4^-$.

Unlike $W_{1/3}$ \eqref{trivial1/3}, the line $W_1^+$ 
\eqref{1/3to1/2} is 1/2 BPS, but if we insert $\sigma$ 
into it as in \eqref{Wsigma}, it is natural 
to analyse it in the same context as the 1/3 BPS line.
In the following we study both objects: the true 
1/3 BPS line constructed from a larger superconection and the 
1/2 BPS line with the topological operator $\sigma$ in its spectrum 
and splitting the 1/2 BPS supermultiplets to 1/3 BPS ones.

A subtlety when writing expressions like \eqref{Wsigma} in terms of ABJM 
fields, is that one should treat the $\sigma$'s as a book-keeping 
device indicating where the connection and spectrum of insertions 
changes. One is not meant to implement a substitution rule like 
$\sigma\bar\psi^1_+\bar\sigma=\bar\psi^4_-$. Such notations may also be 
possible, but they are not what we use here.

Not all line operators are Wilson lines, for example 
the vortex loops of \cite{Drukker:2008jm}, and in those 
cases we may not be able to write the expression on the 
right hand side of \eqref{Wsigma} explicitly. For that 
reason, we try to rely as much as possible on algebra, 
rather than on the explicit realisation of $W_{1/3}$ 
and the operator insertions.

\section{Displacement multiplets of 1/3 BPS line operators}
\label{sec:displacements}

Among all operator insertions into the Wilson loop, the displacement operator and its 
superpartners are special, as they arise from broken global symmetries. 
The conservation equation for translation, supersymmetry 
and R-transformations are violated by the Wilson lines. 

We study here how the displacement multiplets of 1/2 BPS 
line defects constructed 
in~\cite{Bianchi:2017ozk, Bianchi:2018scb,Bianchi:2020hsz} 
split into 1/3 BPS multiplets. Most of the analysis 
is based on the breaking of global symmetries, so valid for 
any 1/3 BPS line operator.

\subsection{First 1/2 BPS line}
\label{sec:W1-mult}

The 1/2 BPS defect along the $x=x_3$ axis 
preserves the rigid 1d conformal group, rotation around the line, 
and an $SU(3)\times U(1)$ R-symmetry, 
rotating $I,J=2,3,4$ indicated by $i,j$. 
In addition, it preserves the supercharges 
$Q^{12}_+$, $Q^{13}_+$, $Q^{14}_+$, 
$Q^{23}_-$, $Q^{34}_-$, $Q^{24}_-$ and the corresponding $S$'s.
A realisation of such an operator is $W_1^+$ with the 
superconnection in \eqref{L1}.

For all lines along $x$, the broken translation generators 
are the components $T^{\mu1}$ and $T^{\mu2}$ of the stress tensor. 
The other symmetries broken by this particular line include%
\footnote{Broken rotations, special conformal transformations and superconformal 
generators all vanish at the origin, so do not give further operaotrs.}
the six components of the 
supercurrents $S^{\mu 1i}_-$ and $S^{\mu ij}_+$ and the 6 components of 
the $R$ current $J^\mu{}\indices{_1^i}$ and $J^\mu{}\indices{_i^1}$.
 The conservation equations for the currents are then
\bal
\label{W1-breaking}
\partial_\mu T^{\mu\nu}(x)W_1^+
&=\delta(x_1)\delta(x_2)\delta^\nu_n\,W_1^+[\bD^n(x)]\,,\qquad n\in\{2,3\}\,,
\\
\partial_\mu S^{\mu IJ}_\alpha(x)W_1^+&=\delta(x_1)\delta(x_2)
\,W_1^+[\delta^I_1\delta_\alpha^-\bar\bLambda^J(x)
+\epsilon^{1IJK}\delta_\alpha^+\bLambda_K(x)]\,,
\\
\partial_\mu J^\mu{}\indices{_I^J}(x)W_1^+&=\delta(x_1)\delta(x_2)
\,W_1^+[\delta_I^1\bO^J(x)+\delta^J_1\bar\bO_I(x)]\,.
\eal
Together, the operators on the right hand side form most of the displacement 
multiplet~\cite{Bianchi:2017ozk, Bianchi:2020hsz} including the displacement 
itself $\bD=\bD^1-i\bD^2$, a fermionic operator $\bLambda_i$ and 
the tilt $\bO^i$. 
In fact there is one element missing, $\bF$, which is fermionic and the lowest 
weight state in the multiplet. There are eight further operators in the complex 
conjugate multiplet.

The action of the preserved supersymmetries on the multiplet are 
\cite{Bianchi:2018scb, Bianchi:2020hsz}
\beq{}
\label{1/2preserved1}
\{Q^{1i}_+,\bF\}=\bO^i\,,\quad
[Q^{1i}_+,\bO^j]=\epsilon^{ijk}\bLambda_k\,,\quad
\{Q^{1i}_+,\bLambda_j\}=-2\delta^i_j\bD\,,\quad
[Q^{1i}_+,\bD]=0\,.
\eeq
From the Jacobi identities for the superalgebra one also finds $[Q^{ij}_-,\bF]=0$ and
\beq{}
\label{1/2preserved2}
%[Q^{ij}_-,\bF]=0\,,\quad
[Q^{ij}_-,\bO^k]=-2i\epsilon^{ijk}\cD_x\bF\,,\quad
\{Q^{ij}_-,\bLambda_k\}=2i\delta^i_k\cD_x\bO^j-2i\delta^j_k\cD_x\bO^i\,,\quad
[Q^{ij}_-,\bD]=i\epsilon^{ijk}\partial_x\bLambda_k\,.
\eeq
$\cD_x$ is an appropriate covariant derivative along the line operator.

Explicit expressions for the operators in terms of the fields of ABJM theory are 
presented in~\cite{Bianchi:2020hsz} and also in Appendix~\ref{app:fields}.
These operators can also be identified with fluctuations of the sigma-model describing an 
$AdS_2$ string in $AdS_4\times\bC\bP^3$ \cite{Correa:2014aga}.

\subsection{Second 1/2 BPS line}
\label{sec:W4-mult}

The second line we consider also preserves the conformal group along 
$x$, rotation around the line, and an $SU(3)\times U(1)$ R-symmetry, 
rotating $I,J=1,2,3$, now indicated as $\hat\imath,\hat\jmath$. 
It preserves the supercharges 
$Q^{12}_+$, $Q^{13}_+$, $Q^{23}_+$, 
$Q^{34}_-$, $Q^{24}_-$, $Q^{14}_-$ and the corresponding $S$'s.
A realisation of such an operator is $W_4^-$ with the 
superconnection in \eqref{L4}.

We can write the action of the symmetries broken by $W_4^-$ on 
that loop as
\bal
\label{W4-breaking}
\partial_\mu T^{\mu\nu}(x)W_4^-
&=\delta(x_1)\delta(x_2)\delta^\nu_n\,W_4^-[\Db^n(x)]\,,\qquad n\in\{2,3\}\,,
\\
\partial_\mu S^{\mu IJ}_\alpha(x)W_4^-&=\delta(x_1)\delta(x_2)
\,W_4^-[\epsilon^{IJK4}\delta_\alpha^-\bar\Lambdab_I(x)
+\delta^J_4\delta_\alpha^+\Lambdab^K(x)]\,,
\\
\partial_\mu J^\mu{}\indices{_I^J}(x)W_4^-&=\delta(x_1)\delta(x_2)
\,W_4^-[\delta_4^J\Ob_I(x)+\delta^4_I\bar\Ob^J(x)]\,.
\eal

These operators fit into the displacement multiplet (and its conjugate) 
for the appropriate $\su(1,1|3)$ superalgebra. Compared to the previous 
case we need to exchange 
$1\leftrightarrow4$, though one should also take into account that the 
spinors change chirality as do some signs in the matrix 
$M\indices{^I_J}$.

The analogue of \eqref{1/2preserved1} is now
\beq{}
\label{W4preserved1}
\{Q^{\hat\imath\hat\jmath}_+,\Fb\}
=\epsilon^{\hat\imath\hat\jmath\hat k}\Ob_{\hat k}\,,\quad
[Q^{\hat\imath\hat\jmath}_+,\Ob_{\hat k}]
=\delta^{\hat\imath}_{\hat k}\Lambdab^{\hat\jmath}
-\delta^{\hat\jmath}_{\hat k}\Lambdab^{\hat \imath}\,,\quad
\{Q^{\hat\imath\hat\jmath}_+,\Lambdab^{\hat k}\}=-2\epsilon^{\hat\imath\hat\jmath\hat k}\Db\,,\quad
[Q^{\hat\imath\hat\jmath}_+,\Db]=0\,,
\eeq
and the analogue of \eqref{1/2preserved2} is
\beq{}
\label{W4preserved2}
%[Q^{i\hat\jmath}_-,\bF]=0\,,\quad
[Q^{\hat\imath4}_-,\Ob_{\hat\jmath}]=-2i\delta^{\hat\imath}_{\hat\jmath}\cD_x\Fb\,,\quad
\{Q^{\hat\imath4}_-,\Lambdab^{\hat\jmath}\}=2i\epsilon^{\hat\imath\hat\jmath\hat k}\cD_x\Ob_{\hat k}\,,\quad
[Q^{\hat\imath4}_-,\Db]=i\cD_x\Lambdab^{\hat\imath}\,.
\eeq

\subsection{Decompsition into 1/3 BPS multiplets}

The 1/2 BPS displacement multiplets are in the representation 
${\mathcal B}^{1/2}_{3/2,0,0}$ of their respective $\su(1,1|3)$ 
in the notations of \cite{Bianchi:2017ozk} and 
$\mathbf{L\bar A_1}$ with primary $[\frac32]_{1/2}^{0,0}$ in the notation 
of~\cite{Agmon:2020pde}. This representation splits into two representations 
of $\su(1,1|2)$ denoted as $\mathbf{L\bar A_1}$ with primaries 
$[\frac12]_{1/2}^{0}$ and $[1]_{1}^{0}$ in the notations of~\cite{Agmon:2020pde}.

A simple way to see this in practice is to match the symmetries 
broken by both $W_1^+$ and $W_4^-$ or only one of them.
The symmetries broken by $W_1^+$ and preserved by $W_4^-$ are
\beq
Q_+^{23}\,,\quad
Q_-^{14}\,,\quad
J\indices{_1^a}\,,\quad
J\indices{_a^1}\,,
\qquad a=2,3\,.
\eeq
Those give rise to the operators
\beq
\label{tilt}
Q^{23}_+\rightsquigarrow\bLambda_4\,,\quad
J\indices{_1^2}\rightsquigarrow\bO^2\,,\quad
J\indices{_1^3}\rightsquigarrow\bO^3\,,\quad
\bF\,,\quad
\eeq
and their complex conjugates. We include $\bF$ to complete the multiplet and 
in the following often 
omit the subscript 4 from the singlet $\bLambda$. We call this the tilt multiplet.

Likewise the symmetries broken by $W_4^-$ and not by $W_1^+$ are
\beq
Q_+^{14}\,,\quad
Q_-^{23}\,,\quad
J\indices{_4^a}\,,\quad
J\indices{_a^4}\,,
\qquad a=2,3\,.
\eeq
Those give rise to the operators
\beq
\label{tlit}
Q^{14}_+\rightsquigarrow\Lambdab^4\,,\quad
J\indices{_2^4}\rightsquigarrow\Ob_2\,,\quad
J\indices{_3^4}\rightsquigarrow\Ob_3\,,\quad
\Fb\,,\quad
\eeq
and the complex conjugate multiplet. We name this the tlit multiplet, 
to distinguish from the tilt.

The symmetries broken by both $W_1^+$ and $W_4^-$ are
\beq
P_1\,,\quad
P_2\,,\quad
Q_-^{12}\,,\quad
Q_-^{13}\,,\quad
Q_+^{24}\,,\quad
Q_+^{34}\,,\quad
J\indices{_1^4}\,,\quad
J\indices{_4^1}\,,
\eeq
In the case of $W_1^+$ they correspond to
\beq
\label{disp}
P_1-iP_2\rightsquigarrow\bD\,,\quad
Q_+^{24} \rightsquigarrow-\bLambda_3\,,\quad
Q_+^{34} \rightsquigarrow\bLambda_2\,,\quad
J\indices{_1^4} \rightsquigarrow\bO^4\,,
\eeq
and in the case of $W_4^-$
\beq
\label{psid}
P_1-iP_2\rightsquigarrow\Db\,,\quad
Q_+^{24} \rightsquigarrow\Lambdab^2\,,\quad
Q_+^{34} \rightsquigarrow\bLambda^3\,,\quad
J\indices{_1^4} \rightsquigarrow\Ob_1\,.
\eeq

\subsection{Multiplets of $W_{1/3}$}
\label{sec:1/3-mult}

In the case of a 1/3 BPS line operator, based on symmetry 
breaking alone, we should have the combination of terms in 
\eqref{W1-breaking} and \eqref{W4-breaking}
\bal
\label{1/3breaking}
\partial_\mu T^{\mu\nu}(x)W_{1/3}
&=\delta(x_1)\delta(x_2)\delta^\nu_n\,W_{1/3}\big[\bD^n(x)\big]\,,\qquad n\in\{2,3\}\,,
\\
\partial_\mu S^{\mu IJ}_\alpha(x)W_{1/3}&=\delta(x_1)\delta(x_2)
\,W_{1/3}\big[
\epsilon^{1aI4}\delta^J_4\delta_\alpha^+\bLambdab_a(x)
+\delta^I_1\delta^J_a\delta_\alpha^-\bar\bLambdab^a(x)
\\&\quad
+\delta^I_2\delta^J_3
\big(\delta_\alpha^+\bLambda(x)+\delta_\alpha^
-\bar\Lambdab(x)\big)
+\delta^I_1\delta^J_4
\big(\delta_\alpha^+\Lambdab(x)+\delta_\alpha^-\bar\bLambda(x)\big)
\big]\,,
\\
\partial_\mu J^\mu{}\indices{_I^J}(x)W_{1/3}&=\delta(x_1)\delta(x_2)
\,W_{1/3}\big[
\delta_I^1\delta^J_4\bOb(x)
+\delta_I^4\delta^J_1\bar\bOb(x)
+\delta_I^1\delta^J_a\bO^a(x)
\\&\quad
+\delta_I^a\delta^J_1\bar\bO_a(x)
+\delta_I^a\delta^J_4\Ob_a(x)
+\delta_I^4\delta^J_a\bar\Ob^a(x)\big]\,.
\eal

For $W_{1/3}$ in \eqref{trivial1/3}, where the connection 
is a larger supermatrix, the operators on the right hand side 
are now matrices. In the $n_1=n_4=1$ case, 
the operators form the tilt multiplet are
\beq
\label{1/3tilt}
\begin{pmatrix}\bF&0\\0&0\end{pmatrix},\quad
\begin{pmatrix}\bO^a&0\\0&0\end{pmatrix},\quad
\begin{pmatrix}\bLambda_4&0\\0&0\end{pmatrix}.
\eeq
The tlit multiplet is
\beq
\label{1/3tlit}
\begin{pmatrix}0&0\\0&\Fb\end{pmatrix},\quad
\begin{pmatrix}0&0\\0&\Ob_a\end{pmatrix},\quad
\begin{pmatrix}0&0\\0&\Lambdab^1\end{pmatrix},
\eeq
and the displacement multiplet is
\beq
\label{1/3disp}
\bOb=\begin{pmatrix}\bO^4&0\\0&\Ob_1\end{pmatrix},\quad
\bLambdab_a=\begin{pmatrix}\bLambda_a&0\\0&\epsilon_{ab}\Lambdab^b\end{pmatrix},\quad
\bDb=\begin{pmatrix}\bD&0\\0&\Db\end{pmatrix}.
\eeq
We do not introduce different notation for the matrices in 
\eqref{1/3tilt} and \eqref{1/3tlit} and at times below refer 
to the entire larger matrices with the same letter as the 
operator inside. It should be clear from the context, 
which of those is meant.

The action of the preserved generators 
on the different operators are presented in Appendix~\ref{app:commutators}. 
Explicit expressions for these operators in terms of the ABJM fields 
are presented in Appendix~\ref{app:tilt-fields}. 

In addition to the multiplets inherited from the 1/2 BPS displacement 
multiplet, we have the permutation multiplet  
constructed in Appendix~\ref{app:sigma-mult}
\beq
\label{1/3sigma}
\sigma=\begin{pmatrix}0&1\\1&0\end{pmatrix},\qquad
\bSigma^a=\begin{pmatrix}0&\bar{G}^a -G^a\\G^a-\bar G^a&0\end{pmatrix},\qquad
o\,.
\eeq
For the expression for $o$, see \eqref{o-def}. We can also think of them 
more abstractly as a short representation of the 1/3 BPS algebra with a primary 
with labels $[0]^0_0$ in the $\mathbf{A_1\bar A_1}$ multiplet 
the notations of~\cite{Agmon:2020pde}. Unlike the fields in the other 
multiplets, this operator is real.

We can of course form composites of these operators, which include 
the combinations like $\bF\bLambda$ as well as off-diagonal 
entries arising from $\sigma$ times another operator. As usual, if two 
operators do not share supercharges, the composite would not be protected. 
An example of that is $\sigma$ and $\bOb$.

We can also endow the operators with Chan-Paton factors
\beq
\Fb\indices{_1^1}\simeq\begin{pmatrix}\Fb&0\\0&0\end{pmatrix},\quad
\Fb\indices{_1^2}\simeq\begin{pmatrix}0&\Fb\\0&0\end{pmatrix},\quad
\Fb\indices{_2^1}\simeq\begin{pmatrix}0&0\\\Fb&0\end{pmatrix},\quad
\Fb\indices{_2^2}\simeq\begin{pmatrix}0&0\\0&\Fb\end{pmatrix}.
\eeq
In particular operators like $\Fb\indices{_1^1}=\sigma\Fb\indices{_2^2}\sigma$ 
are inserted into the $\cL_1^+$ line and 
enable our construction in the next subsection.

\subsection{1/3 BPS multiplets of the 1/2 BPS Wilson line}

The 1/2 BPS line $W_1^+$ has a displacement multiplet, as presented 
in Section~\ref{sec:W1-mult}. It combines the 
1/3 BPS displacement multiplet with $\bO^4$, $\bLambda_a$ and $\bD$ 
as well as the 1/3 BPS tilt with $\bF$, $\bO^a$ and $\bLambda$ 
\eqref{tilt}.

As presented in Section~\ref{sec:1/2sigma}, 
The operators $\sigma^+$ and $\sigma^-$ (or $\sigma$ and $\bar\sigma$) 
\eqref{sigmapm} are also natural insertions in $W_1^+$. 
They are in the 1/3 BPS multiplet \eqref{1/3sigma} and as $\sigma$ 
changes the superconnection from $\cL_1^+$ to $\cL_4^-$, we 
can then insert operators naturally living on $W_4^-$. 
In particular this applies to the tlit operators, \eqref{tlit} as
\beq
\sigma\Fb\,,\qquad
\sigma\Ob_a\,,\qquad
\sigma\Lambdab\,.
\eeq
We can adjoin to all these operators $\bar\sigma$ from the right, so 
they become insertions into the $W_1^+$ loop without a change in 
connection.

Some care is required in analysing $\sigma$ and composites like 
$\sigma\Fb$ or $\sigma\Fb\bar\sigma$. Recall that a special 
feature of the ABJM Wilson lines is that the preserved supercharges 
do not annihilate the connections in Appendix~\ref{app:WLs} 
but give total derivatives \eqref{total-deriv}. When acting on the entire line, 
these total derivative integrate to zero, hence the loops are BPS. 
This is rather similar to the insertion of BPS protected operator 
into the BPS cusp in $\cN=4$ SYM in 4d~\cite{Gromov:2012eu}.

When acting on the line with insertions, we find extra boundary 
terms. As discussed in Appendix~\ref{app:subtlety}, the action 
of a supercharge on 
an odd supermatrix insertion in $W_1^-$ is covariantised to
\bal
\label{covariantisation}
\tilde Q_+^{1a} \bullet =Q_+^{1a} \bullet - \{\bar G^a, \bullet\}\,,
\qquad 
\tilde Q_-^{a4} \bullet =Q_-^{a4} \bullet -\{ G^a , \bullet\}\,,
\eal
with $\bar G^a$ and $G^a$ in \eqref{G^a}. In the case of $W_4^-$ the 
roles of $\bar G^a$ and $G^a$ are reversed \eqref{total-deriv}.

In evaluating the variation of $\sigma$, the direct action 
by $Q_\alpha^{IJ}$ is trivial and we only have the covariant 
part, with that from $W_1^+$ on the left and from $W_4^-$ on the 
right. This is the source of the terms in the expression for 
$\bSigma$ \eqref{1/3sigma}. 

An $\Fb$ insertion into $W_4^-$ is annihilated by three 
supercharges of which two are shared by $W_1^+$. Yet, 
when inserting it as $\sigma\Fb\bar\sigma$ into $W_1^+$, 
there are different total derivative terms. In fact, 
no supercharges annihilate 
it and only the combination of $Q^{1a}_++Q^{a4}_-$ acting 
on it gives $\epsilon^{ab}\sigma\Ob_b\bar\sigma$.

This construction seems to introduces several new marginal operators 
into the 1/2 BPS loop: $\sigma\Ob_a\bar\sigma$, $o\bar\sigma$ and 
$(\cD_x\sigma)\bar\sigma$ \eqref{Dsigma}. Unlike $\bO^2$, $\bO^3$ and 
$\bO^4$, these operators do not arise from broken global symmetries, 
so it is not guaranteed that they are indeed exactly marginal. 
We leave this question for further study.

Of particular note is the operator $(\cD_x\sigma)\bar\sigma$, the 
descendant of $\sigma$ \eqref{Dsigma}, which is an infinitesimal 
deformation in the direction of the ABJM version of the loops described 
in Section~6.3.2 of~\cite{Drukker:2022ywj}. The loops described 
there are classically conformal, but conformality is not guaranteed by 
the preserved supercharges. The question raised in the last paragraph 
is another avatar of the question of whether these loops are truly 
conformal.

For explicit expressions in terms of the ABJM fields, see 
Appendix~\ref{app:tilt-fields}.

\section{Two point functions}
\label{sec:2pt}

For the operators arising from broken symmetries, as in \eqref{sec:1/3-mult}, 
their normalisations are fixed by the normalisation of the conserved currents. 
We study here the relations between the normalisations of the different 
operators and their relation to the bremsstrahlung functions of these 
loops.

\subsection{Ward identities}
\label{sec:ward}

From conformal symmetry we know that the correlators of the operators in the displacement 
multiplet take the form
\begin{align}
\label{CD}
\llangle\bDb(0)\bar\bDb(x)\rrangle&=\frac{C_{\bDb}}{x^4}\,,\\
\llangle\bLambdab_a(0)\bar\bLambdab^b(x)\rrangle
&=\frac{C_{\bLambdab_a}\delta_a^b}{x^3}\,,
\\
\label{CO}
\llangle\bOb(0)\bar\bOb(x)\rrangle&=\frac{C_{\bOb}}{x^2}\,.
\end{align}
The notation $\llangle\bullet\dots\bullet\rrangle$ represents the 
expectation value of the $\bullet$ insertions into the line 
normalized by the expectation value of the line without insertions.

The coefficients $C_\bDb$ and $C_\bOb$ are fixed from the definition of the operators and the 
normalisation of the broken currents in \eqref{1/3breaking}. 
They are also related to the bremsstrahlung functions 
of the line operators, as in \eqref{D-B}, \eqref{O-B}. The relations between them can be 
found from the ward identity for supersymmetry, 
following~\cite{Bianchi:2018scb, Bianchi:2018zpb}.

Starting with the vanishing correlator $\llangle\bLambdab_2(0)\,\bar\bDb(x)\rrangle=0$ 
and acting with the preserved supercharge $Q^{12}_+$, using \eqref{1/2preserved1}, we find
\beq
\label{bdbbLb}
-2\llangle\bDb(0)\,\bar\bDb(x)\rrangle
=\llangle\bLambdab_2(0)\,\cD_x\bar\bLambdab^2(x)\rrangle\,,
\eeq
where $\cD_x$ is an appropriate covariant derivative along the Wilson line. This gives 
$2C_\bDb=3C_{\bLambdab_a}$. 

Likewise starting with $\llangle\bOb(0)\,\bar\bLambdab^3(x)\rrangle=0$
and acting with the preserved supercharge $Q^{12}_+$ as in \eqref{1/2preserved2}, we find
\beq
\label{bLbbOb}
-\llangle\bLambdab_3(0)\,\bar\bLambdab^3(x)\rrangle
=2\llangle\bOb\,\cD_x\bar\bOb(x)\rrangle\,,
\eeq
or $C_{\bLambdab_a}=4C_{\bOb}$. Combining the two, we find $C_\bDb=6C_\bOb$. 

These expressions were already derived in~\cite{Bianchi:2020hsz} from a superspace 
representation of the displacement multiplet in the case of the 1/2 BPS loop and they 
are not modified in the 1/3 BPS case. 

For the operators in the tilt and tlit multiplets
\begin{align}
%\label{CD}
\llangle\bLambda(0)\bar\bLambda(x)\rrangle&=\frac{C_{\bLambda}}{x^3}\,,
&
\llangle\Lambdab(0)\,
\bar\Lambdab(x)\rrangle&=\frac{C_{\Lambdab}}{x^3}\,,
\\
\label{CbO}
\llangle\bO^a(0)\bar\bO_b(x)\rrangle&=\frac{C_{\bO^a}\delta_b^a}{x^2}\,,
&
\llangle\Ob_a(0)\,
\bar\Ob^b(x)\rrangle&=\frac{C_{\Ob_a}\delta^b_a}{x^3}\,,
\\
\llangle\bF(0)\bar\bF(x)\rrangle&=\frac{C_{\bF}}{x}\,.
&
\llangle\Fb(0)\,
\bar\Fb(x)\rrangle&=\frac{C_{\Fb}}{x^3}\,.
\end{align}
To find relations among those, we start with the vanishing correlator 
$\llangle\bO^3(0)\,\bar\bLambda(x)\rrangle$ and act with 
the preserved supercharge $Q^{12}_+$, which yields
\beq
-\llangle\bLambda(0)\,\bar\bLambda(x)\rrangle
=2\llangle\bO^3(0)\,\cD_x\bar\bO_3(x)\rrangle
\quad\Rightarrow\quad
C_\bLambda=4C_{\bO^a}\,.
\eeq
Then taking $\llangle\bF(0)\,\bar\bO_2(x)\rrangle$ and acting with 
the preserved supercharge $Q^{12}_+$, we find
\beq
-\llangle\bO(0)\,\bar\bO(x)\rrangle
=2\llangle\bF^2(0)\,\cD_x\bar\bF_2(x)\rrangle\,,
\eeq
This gives $C_{\bO^a}=2C_{\bF}$ and finally $C_{\bLambda}=8C_{\bF}$.

The expressions for the tlit multiplet are identical, but 
$C_{\bO^a}$ does not have to be equal to $C_{\Ob^a}$. 
Likewise, for the 1/2 BPS loop we know that $C_{\bLambda}=C_{\bLambdab_a}$ and 
$C_{\bO^a}=C_{\bOb}$, but this does not necessarily hold for 1/3 BPS operators, 
as we discuss in the next section.

\subsection{Relations accross multiplets}
\label{sec:broken}

We can go further and relate the different multiplets to each-other, using 
the explicit representation of $W_{1/3}$ \eqref{trivial1/3} and its expression 
in terms of 1/2 BPS loops. We consider 
the case of $n_1$ copies of $\cL_1^+$ and $n_4$ copies of $\cL_4^-$, but 
for simplicity write them as $2\times2$ matrices.

Using the representation in \eqref{1/3tilt}, the two point function of 
the operators from the tilt multiplets can be related to those of the 
1/2 BPS loop $W_1^+$ as
\beq
\label{CbO1/3-1/2}
C^{1/3}_{\bO^a}\delta_b^a
=x^2\LLangle\begin{pmatrix}\bO^a (0) &0\\0& 0\end{pmatrix}
\begin{pmatrix}\bar{\bO}_b(x) & 0\\0& 0\end{pmatrix}
\RRangle
=\frac{n_1\delta^a_b}{n_1+n_4}C_\bO^{1/2}\,.
\eeq
This is a simple consequence of having $n_1$ insertions. Likewise 
for $\Ob_a$ and $\bOb$, we have
\beq
\label{C1/3-1/2}
C^{1/3}_{\Ob_a}=\frac{n_4}{n_1+n_4}C_\bO^{1/2}\,,
\qquad
C^{1/3}_{\bOb}=C_\bO^{1/2}\,.
\eeq
In particular
\beq
\label{broken-ward}
C^{1/3}_{\bO^a}+C^{1/3}_{\Ob_a}=C^{1/3}_{\bOb}\,.
\eeq
Though we derived this from the expressions in \eqref{trivial1/3}, we 
expect this relation to hold for any 1/3 BPS loop.

The story is very different when considering the operators inserted into 
the 1/2 BPS Wilson loop with the aid of $\sigma$. 
In that case $C_{\bO^a}=C_{\bO^4}=C_\bO^{1/2}$, as 
these are simply the usual tilt operators of the 1/2 BPS line. If we 
look at 
$\llangle\sigma\Ob_a\bar\sigma(0)\,\sigma\bar\Ob^b\bar\sigma(x)\rrangle$, 
assume $C_\sigma=1$ to cancel the middle $\sigma(0)\bar\sigma(x)$ and 
move the other $\sigma$'s far away, then it 
is natural to expect that this too is $C_{\bO}^{1/2}$. 
This indicates that in this case all three normalisation constants are 
equal to $C_\bO^{1/2}$, though this deserves more careful study.

\subsection{Bremsstrahlung functions of $W_{1/3}$}
\label{sec:brems}

As reviewed in Appendix~\ref{app:CBD}, the normalisation constants are 
related to the bremsstrahlung functions arising from nearly straight 
cusps.

We can characterise cusps of the 1/3 BPS loop \eqref{trivial1/3} by 
an angle $\phi$ and an $R$ symmetry $SU(4)$ matrix $U$. 
When this matrix is close to the identity we 
can write it in terms of the broken symmetry generators as
\beq
U=I+i\theta J\indices{_1^4}
+i\theta_a'J\indices{_1^a}
+i\theta^{\prime\prime b}J\indices{_b^4}\,.
\eeq
Then the cusp anomalous dimension takes the form (where we omit the indices 
from the $\theta$)
\beq
\label{general-cusp}
\Gamma(\phi,U)\simeq 
B^{1/3}_\theta\theta^2+\theta'^2B^{1/3}_{\theta'}+\theta''^2B^{1/3}_{\theta''}
-\phi^2B^{1/3}_\phi\,.
\eeq
$W_{1/3}$ has therefore four brensstrahlung functions and the usual 
relations \eqref{D-B} and \eqref{O-B} give
\beq
B^{1/3}_\phi=\frac{1}{24}C^{1/3}_\bDb\,,
\quad
B^{1/3}_\theta=\frac{1}{4}C^{1/3}_\bOb\,,
\quad
B^{1/3}_{\theta'}=\frac{1}{4}C^{1/3}_{\bO^a}\,,
\quad
B^{1/3}_{\theta''}=\frac{1}{4}C^{1/3}_{\Ob_a}\,.
\eeq

The relation after \eqref{bLbbOb} and \eqref{broken-ward} then lead to the 
equalities
\beq
B^{1/3}_\phi
=B^{1/3}_\theta
=B^{1/3}_{\theta'}
+B^{1/3}_{\theta''}\,.
\eeq
This allows us to write \eqref{general-cusp} in terms of only two 
independent functions
\beq
\Gamma(\phi,U)\simeq 
(\theta^2+\theta'^2-\phi^2)B^{1/3}_{\theta'}
+(\theta^2+\theta''^2-\phi^2)B^{1/3}_{\theta''}\,.
\eeq
Furthermore, we can rely on the relation to the 1/2 BPS loop \eqref{C1/3-1/2} 
to write this in terms of the 1/2 BPS bremsstrahlung function $B_\phi^{1/2}$ 
and $n_1$, $n_4$ as
\beq
\label{final-brem}
\Gamma(\phi,U)\simeq 
\left(\theta^2-\phi^2+\frac{n_1}{n_1+n_4}\theta'^2
+\frac{n_4}{n_1+n_4}\theta''^2\right)B_\phi^{1/2}\,.
\eeq

This relation can be seen as a generalisation of that found for the 
1/6 BPS bosonic loop, where 
$2B_\theta^\text{bos}=B_\phi^\text{bos}$~\cite{Bianchi:2017svd, 
Bianchi:2017ozk}. 
To see the relation, take $\theta=\theta''=0$ and $n_1=n_4$ in 
\eqref{final-brem} and then identify $\theta'$ with the angle in 
the 1/6 BPS cusp.

\section{Defect conformal manifolds}
\label{sec:DCM}

We identified multiple marginal operators living on the 1/3 BPS line 
as well as possible new marginal operators on the 1/2 BPS line. Such operators 
allow to deform the defect along a defect conformal manifold, the space 
of all connected conformal defects. For complex marginal operators 
$\Phi_i$ and $\bar\Phi_{\bar\imath}$ we define the coordinates 
$\zeta^i$ and $\bar\zeta^{\bar\imath}$ and express the deformed line as
\beq
\label{deformedW}
W_{\zeta,\bar\zeta}[\bullet\dots\bullet]=
W\left[\bullet\dots\bullet\,
\exp\int dx\left(\zeta^i \Phi_i(x)
+\bar\zeta^{\bar\imath}\bar \Phi_{\bar\imath}(x)\right)\right].
\eeq

This space of dCFTs is endowed with the Zamolodchikov metric
\beq
\label{dZam}
g_{i\bar\jmath}(\zeta,\bar\zeta)
=\frac{\langle W_{\zeta,\bar\zeta}[\Phi_i(0)\bar \Phi_{\bar\jmath}(\infty)]\rangle}
{\langle W_{\zeta,\bar\zeta}\rangle}\,,
\qquad
\bar \Phi_{\bar\jmath}(\infty)=\lim_{x\to\infty}x^2\bar \Phi_{\bar\jmath}(x)\,.
\eeq
Clearly at $\zeta=\bar\zeta=0$ the metric is given by expressions like 
$C_{\bO^a}$, $C_{\Ob_a}$ and $C_{\bOb}$ \eqref{CO}, \eqref{CbO}. According 
to~\cite{Kutasov:1988xb,Friedan:2012hi} the curvature is as in 
Riemann normal coordinates; the second derivative of the metric with 
respect to the coordinates, leading to the integrated 4-point function
\bal
\label{friedan1d}
R_{i\bar\jmath k\bar l}&=\int_{-\infty}^{+\infty}dx_1 \,dx_2
\Big(\!\llangle \bar{\Phi}_{\bar{\jmath}} (x_1) \Phi_k (x_2) \Phi_i (0) 
\bar{\Phi}_{\bar{l}} (\infty) \rrangle_c 
-\llangle \Phi_{i} (x_1) \Phi_k (x_2) \bar{\Phi}_{\bar{\jmath}} (0) 
\bar{\Phi}_{\bar{l}} (\infty) \rrangle_c\!\Big) \\
&=-\text{RV}\int_{-\infty}^{+\infty} d\eta \log|\eta|
\Big(\!\llangle\Phi_i (1) \bar\Phi_{\bar\jmath} (\eta) 
\Phi_k (\infty) \bar\Phi_{\bar l}(0)\rrangle_c 
+ \llangle\Phi_i (0)\bar\Phi_{\bar\jmath}(1-\eta) \Phi_k(\infty) \bar \Phi_{\bar l}(1)\rrangle_c\!
\Big)\,,
\eal
and likewise for other components of the curvature. 
The subscript $c$ indicates the connected component, RV is a regularisation 
prescription detailed in~\cite{Friedan:2012hi} and the order of the operators 
is not as indicated in the second line, but should 
be by increasing argument which depends on the value of 
$x_1$ and $x_2$ or $\eta$. Explicitly
\bal
\llangle \bar{\Phi}_{\bar{\jmath}} (x_1) \Phi_k (x_2) \Phi_i (0) 
\bar{\Phi}_{\bar{l}} (\infty) \rrangle_c=\begin{cases}
\llangle \bar{\Phi}_{\bar{\jmath}} (x_1) \Phi_k (x_2) \Phi_i (0) 
\bar{\Phi}_{\bar{l}} (\infty) \rrangle_c& 
\text{for } x_1<x_2<0\,,
\\
\llangle \bar{\Phi}_{\bar{\jmath}} (x_1) \Phi_i (0) \Phi_k (x_2)
\bar{\Phi}_{\bar{l}} (\infty) \rrangle_c&
\text{for } x_1<0<x_2\,,
\\
\llangle \Phi_i (0) \bar{\Phi}_{ \bar{\jmath}} (x_1) \Phi_k (x_2)
\bar{\Phi}_{\bar{l}} (\infty) \rrangle_c&
\text{for } 0<x_1<x_2\,,
\\
\text{other three similar cases for $x_2<x_1$.\hskip-1.2cm}
\end{cases}
\eal

Equation \eqref{friedan1d} was further simplified in~\cite{Drukker:2022pxk}, 
by using crossing symmetry. See the expressions there.

\subsection{The case of $W_{1/3}$}
\label{sec:1/3DCM}

Of all the marginal operators, the simplest ones are those that 
arise from global symmetry breaking. In the case of 
1/3 BPS line operators, they break the global symmetry group $OSp(6|4)$ to 
$SU(1,1|2)\times U(1)\times U(1)$. The $SU(4)$ R-symmetry group is broken to 
$SU(2)\times U(1)\times U(1)$. This indicates that the space of allowed 
1/3 BPS loops is (at least) the quotient
\beq
\label{coset}
\cM=SU(4)/S(U(2)\times U(1)\times U(1))\,.
\eeq
This is a 10 dimensional manifold (or 5 complex-dimensional).

Symmetry breaking gives rise to the tilt, tlit and displacement multiplets 
and they contain five complex operators of dimension one, 
$\bO^a$, $\Ob_a$ and $\bOb$. To conform with the notation 
in \eqref{deformedW}, we label the marginal operators collectively as
\beq
\Phi_i\simeq\{\bO^2,\bO^3,\bOb, \Ob_2,\Ob_3\}\,,\qquad
\bar\Phi_{\bar\imath}\simeq\{\bar\bO_2,\bar\bO_3,\bar\bOb, 
\bar\Ob^2,\bar\Ob^3\}\,,
\qquad
i,\bar\imath=1,\cdots,5\,.
\eeq
For finite $\zeta^1$, $\zeta^2$, the $\cL_1^+$ entries in the 
line \eqref{trivial1/3} are rotated into another one with 
$\cL_{1'}^+$. Finite 
$\zeta^4$ and $\zeta^5$ change the $\cL_4^+$ block. 

The nonvanishing components of the metric are \eqref{CbO1/3-1/2}, \eqref{C1/3-1/2}
\bal
g_{i\bar{\jmath}}=\begin{cases}
C_{\bO^a}^{1/3} \delta_{i\bar\jmath}
=\frac{n_1}{n_1+n_4}C_{\bO}^{1/2} \delta_{i\bar\jmath}\,,
& i,\bar\jmath=1,2\,.\\
C_{\bOb}^{1/3}
=C_{\bO}^{1/2}\,,
& i,\bar\jmath=3\,,\\
C_{\Ob_a}^{1/3} \delta_{i\bar\jmath}
=\frac{n_4}{n_1+n_4}C_{\bO}^{1/2} \delta_{i\bar\jmath}\,,
& i,\bar\jmath=4,5\,.
\end{cases}
\eal
To calculate the curvature we use \eqref{friedan1d}, where we insert 
the operators \eqref{1/3tilt}, \eqref{1/3tlit} 
and \eqref{1/3disp} into the superconnection.
For example, for $i=k=1$ and $\bar\imath=\bar l=1$
\bal
\label{simple-4}
R_{1\bar{1} 1 \bar{1}}=\int_{-\infty}^{+\infty}
dx_1\, dx_2 
&\bigg[\LLangle \begin{pmatrix}\bar{\bO}_2(x_1) & 0\\0&0\end{pmatrix}
\begin{pmatrix}\bO^2(x_2)  & 0\\0 & 0\end{pmatrix} 
\begin{pmatrix}\bO^2(0)  & 0\\0&0\end{pmatrix}
\begin{pmatrix}\bar{\bO}_2(\infty) & 0\\0& 0\end{pmatrix} \RRangle_c \\
&-\LLangle \begin{pmatrix}\bO^2(x_1) & 0\\0&0\end{pmatrix}
\begin{pmatrix}\bO^2(x_2)  & 0\\0 & 0\end{pmatrix} 
\begin{pmatrix}\bar\bO_2(0)  & 0\\0&0\end{pmatrix}
\begin{pmatrix}\bar{\bO}_2(\infty) & 0\\0& 0\end{pmatrix} \RRangle_c
\Bigg]\,.
\eal
We write here $2\times2$ matrices, but they should be larger, as 
appropriate. 
This expression involves only the insertions of the tilt operators 
of $W_1^+$ into $W_1^+$, so is the same as in~\cite{Drukker:2022pxk}, 
except for the normalisation, which is $n_1/(n_1+n_4)$, because 
there are no insertions into the $W_4^-$ block. 
The 4-point function in the case of the 1/2 BPS Wilson loop 
was calculated in~\cite{Bianchi:2020hsz} and the integral 
was evaluated in~\cite{Drukker:2022pxk} with the final expression 
(accounting for the normalisation) being
\beq
R_{1\bar{1} 1 \bar{1}}=2g_{1\bar1}=2C^{1/3}_{\bO^a}
=\frac{2n_1}{n_1+n_4}C_{\bO}^{1/2}\,.
\eeq

In Appendix~\ref{app:coset} we calculate the Riemann tensor of 
the quotient \eqref{coset} in a matching coordinate system and 
write down all its nonzero components. Indeed 
$R_{1\bar{1} 1 \bar{1}}=2g_{1\bar1}$, as in the CFT calculation 
above. In the same way we can match all the components of the curvature 
for $c=a+b$ \eqref{explicit-R1} except for terms mixing $1,2$ and $4,5$ 
indices, such as $R_{1\bar44\bar1}$ and $R_{1\bar24\bar5}$.

In those cases, plugging the expressions from \eqref{1/3tilt} and 
\eqref{1/3tlit} into \eqref{friedan1d} would give something like 
\eqref{simple-4}, but with two non-zero entries at the top left and 
two on the bottom right, which seems to vanish.

To fix that, we need another ingredient ignored so far.%
\footnote{We thank V. Schomerus for clarifying this point to us.}
The expression for the tilt and tlit in \eqref{1/3tilt}, \eqref{1/3tlit} 
are the terms arising from symmetry breaking, as in \eqref{1/3breaking}. 
If symmetries are not broken, then there should be a conserved 
current along the line. In the case of the preserved supercharges 
these are the total derivatives in \eqref{total-deriv}, where 
$\bar G^a$ and $G^a$ can be considered as supercurrents along the 
line. For the R-symmetry charges
\beq
[J\indices{_1^a},W_{1/3}]
=\int dx\,W_{1/3}\left[\begin{pmatrix}
     \bO^a & \partial_{x} \Gamma\indices{_1^a}\\
     \partial_{x} \Gamma\indices{_1^a} & \partial_{x} \Gamma\indices{_1^a}
\end{pmatrix}(x)\right],
\eeq
where $\Gamma\indices{_1^a}$ are $SU(4)$ generators and serve as 1d conserved 
currents (they should be written as $\Gamma\indices{^x_1^a}$, but we omit the 
repetitive superscript). Their derivative vanishes, since this symmetry is 
preserved for those three entries, but these expressions are important 
to reproduce the missing components of the curvature.

Looking at the first term in the first line of \eqref{friedan1d} in the 
case of $R_{1\bar44\bar1}$, we get the integrand
\beq
\LLangle \begin{pmatrix}
     \partial_{x} \Gamma\indices{_2^4} & \partial_{x} \Gamma\indices{_2^4}\\
     \partial_{x} \Gamma\indices{_2^4} & \Ob_2
\end{pmatrix} (x_1) 
\begin{pmatrix}
     \partial_{x} \Gamma\indices{_4^2} & \partial_{x} \Gamma\indices{_4^2}\\
     \partial_{x} \Gamma\indices{_4^2} & \bar{\Ob}^2
\end{pmatrix} (x_2) 
\begin{pmatrix}
     \bO^2 & \partial_{x} \Gamma\indices{_1^2}\\
     \partial_{x} \Gamma\indices{_1^2} & \partial_{x} \Gamma\indices{_1^2}
\end{pmatrix}(0) 
\begin{pmatrix}
     \bar{\bO}_2 & \partial_{x} \Gamma\indices{_2^1}\\
     \partial_{x} \Gamma\indices{_2^1} & \partial_{x} \Gamma\indices{_2^1}
\end{pmatrix} (\infty) \RRangle_c.
\eeq
The derivatives $\partial_x\Gamma$ vanish at the points $0$ 
and $\infty$, so we keep there only $\bO^2$ and $\bar\bO_2$. 
We then ignore $\Ob_2$ and $\bar\Ob^2$ from the first two 
terms, since they give disconnected contributions. 
Integration over the remaining $\partial_x\Gamma$, for 
the ordering $0<x_1<x_2$ gives
\bal{}
&{}-\LLangle
\begin{pmatrix}\bO^2 & 0\\0 &0\end{pmatrix}(0) 
\begin{pmatrix}
\Gamma\indices{_2^4} &\Gamma\indices{_2^4}\\
\Gamma\indices{_2^4} & 0
\end{pmatrix} (0) 
\begin{pmatrix}
\Gamma\indices{_4^2} & \Gamma\indices{_4^2}\\
\Gamma\indices{_4^2} & 0
\end{pmatrix} (\infty) 
\begin{pmatrix}\bar{\bO}_2 & 0\\0 & 0\end{pmatrix} (\infty) 
\RRangle_c
\\
&=-\LLangle
\begin{pmatrix}
\bO^2\Gamma\indices{_2^4} &\bO^2\Gamma\indices{_2^4}\\
0& 0\end{pmatrix} (0) 
\begin{pmatrix}
\Gamma\indices{_4^2}\bar{\bO}_2& 0\\
\Gamma\indices{_4^2}\bar{\bO}_2 & 0
\end{pmatrix} (\infty) 
\RRangle_c
\sim-\LLangle
\begin{pmatrix}\bO^4(0)&0\\0& 0\end{pmatrix} 
\begin{pmatrix}\bar{\bO}_4(\infty)& 0\\0& 0\end{pmatrix}
\RRangle.
\eal
It is natural to consider only the off-diagonal terms as 
contributing to the connected part of the correlator, as 
the other terms would arise also in the case of the 1/2 BPS 
loop in \cite{Drukker:2022pxk}.

For a general $W_{1/3}=n_1W_1^++n_4W_4^-$, we would get 
contributions from $n_1n_4$ off-diagonal entries, giving 
the answer
\beq
\label{1441}
R_{1\bar44\bar1}=-n_4C_{\bO^a}^{1/3}=-n_1C_{\Ob_a}^{1/3}
=-\frac{n_1n_4}{n_1+n_4}C_\bO^{1/2}\,,
\eeq
in agreement with \eqref{explicit-R1}. One would expect another 
contribution from the rotations of $\Ob_2$ and $\bar\Ob^2$, but 
one can see that there is no such term in \eqref{friedan1d}. In 
that expression, symmetry was used to reduce four terms to two, 
so we could recover the other contribution and divide them both 
by 2. In any case, they are identical, so the expression in 
\eqref{1441} is correct.

Another case is $R_{1\bar{5}4 \bar{2}}$, where the same calculation yields
\bal
{}&-\LLangle
\begin{pmatrix}
\bO^2\Gamma\indices{_2^4} &\bO^2\Gamma\indices{_2^4}\\
0& 0\end{pmatrix} (0) 
\begin{pmatrix}
\Gamma\indices{_4^3}\bar{\bO}_3 & 0\\
\Gamma\indices{_4^3}\bar{\bO}_3 & 0
\end{pmatrix} (\infty) 
\RRangle_c.
\eal
with the same result as in \eqref{1441}, in agreement with 
\eqref{explicit-R1}. Terms like $R_{1\bar{5}5 \bar{1}}$ 
vanish in \eqref{explicit-R1} and 
this is true also from the field theory side, since 
$\Gamma\indices{_4^3}$ does not act on $\bar\bO_2$. It is 
easy to verify then that such arguments exactly reproduces 
all terms in \eqref{explicit-R1}. 

The results presented above are for the marginal operators 
arising from broken global symmetries. Those are guaranteed to 
be marginal. We mentioned above possible other marginal operators, 
like insertions of $\sigma\Ob_a\bar\sigma$ in $W_1^+$ or $o$ 
and $(\cD_x\sigma)$ from the $\sigma$ multiplet in $W_{1/3}$ 
or for $n_1>1$ an insertion of $\bO_a$ into only one of the 
$\cL_1^+$ blocks. We postpone the question of whether they are 
exactly marginal as well as the resulting conformal manifolds 
to future work.

\section{Discussion}
\label{sec:discuss}

We found an explicit realisation of a 1/3 BPS Wilson line operator 
in ABJM theory in terms of a large superconnection, combining 
two 1/2 BPS Wilson lines, and discussed general properties of 
1/3 BPS line operators. Many of these results are valid for 
any 1/3 BPS loop, including the vortex loop of~\cite{Drukker:2008jm}. 
We have not attempted to verify them by detailed microscopic 
calculations in that setting, as the explicit
forms of defect operators on the vortex loop may be subtle, 
given that there is a singularity along the line. 

The entire discussion was for the straight line operator, but it carries over 
to the case of the circle. The preserved and broken symmetries are related by 
conjugation and we do not think that there are any subtleties in our calculation 
due to the difference between compact and non-compact loops. Of course, when 
we consider only one $\sigma$ insertion in $W_1^+$, we should remember the 
$\bar\sigma$ at infinity, when mapping to the circle.

The circular Wilson loop has a finite expectation value that can be 
calculated using localization~\cite{pestun, Kapustin:2009kz, 
Marino:2009jd, Drukker:2010nc}. Given that $W_{1/3}=n_1W_1^++n_4^-W_4^-$, 
the expression for the 1/3 BPS loop is exactly the same as the 1/2 
BPS one.

The operators $\sigma$, $\bar\sigma$ are a side product of our construction and 
should be studied more fully on their own right. They are presented in 
Section~\ref{sec:permutation}, Section~\ref{sec:1/2sigma} and 
Appendix~\ref{app:sigma-mult}. There is also the $\tau$ operator presented 
under \eqref{sigma} and 
$T$ (studied already in~\cite{Gorini:2022jws}) and the relations among 
them should be examined more closely. For example, whether anything changes 
if we replace $\sigma$ by $\sigma\tau$. With those operators under control, 
one could then try to study operators like $\sigma\Ob_a\bar\sigma$ and 
whether they are truly marginal.

Our analysis of the relation between the normalisation constants in 
Sections~\ref{sec:ward} and~\ref{sec:broken} 
is modeled closely after~\cite{Bianchi:2018scb}. There this was 
done for the 1/6 BPS bosonic Wilson loop, preserving the superalgebra 
$\su(1,1|1)$ and in addition a bosonic $\su(2)\oplus\su(2)$. 
The supermultiplets are much shorter, one has the complex 
displacement $\bDb$ and a superpartner $\bLambdab$ (and 
their descendents). There are also four complex twist 
operators in the $\mathbf{(2,\bar2)}$ representation 
and each has a superpartner.

In that case, the symmetry guaranties that the normalisation factors of all 
of the displacements $C_\bO$ are equal, and a similar result to 
Section~\ref{sec:broken} shows that they are half what they would be if 
they were in the same multiplet as $\bDb$ (as below \eqref{bLbbOb}), 
so $C_\bD=12C_\bO$ and the two bremsstrahlung functions are 
related by this factor of 2.

The defect conformal manifold constructed in Section~\ref{sec:DCM} is a 
generalisation of that in \cite{Drukker:2022pxk}. It is higher dimensional 
and not a symmetric space. Technically we also had to take care of the 
seemingly vanishing mixed curvature terms, which required the inclusion 
of the conserved R-symmetry currents on the line.

For the bosonic loops and their four tilts, the conformal manifold is 
four complex dimensional, and should be 
$SU(4)/S(U(2)\times U(2))=\mathbb{Gr}_2(\bC^4)$, the 
Grassmannian for complex 2-planes in $\bC^4$. Since the preserved 
symmetry includes the $S(U(2)\times U(1)\times U(1))$ of the 1/3 BPS 
loops studied here, our conformal manifold is a $\bC\bP^1$ bundle 
over this Grassmannian. Shrinking the fibers would give the base, in the same 
way we can reduce our conformal manifold in Section~\ref{sec:1/3DCM} 
to that of the 1/2 BPS loop, 
$\bC\bP^3$, by simply taking $n_1\to0$ or $n_4\to0$.

In the case of our 5 complex dimensional conformal manifold the size of 
the $\bC\bP^1$, which is fixed by $C_\bOb$ is related to the other two 
length scales via $C_\bOb=C_{\bO^a}+C_{\Ob_a}$ \eqref{broken-ward}, 
so it cannot be shrunk, without also shrinking the base.

Interestingly, this shrinking can be realised with the aid of the 1/6 BPS 
fermionic loops~\cite{Ouyang:2015iza, Drukker:2019bev}. 
They all preserve the same $\su(1,1|1)$ superalgebra of the bosonic 
loop, but the bosonic symmetry is only 
$SU(2)\times U(1)\times U(1)$, enhancing to $SU(3)\times U(1)$ at the 
1/2 BPS points and $S(U(2)\times U(2))$ at the bosonic point.

The general 1/6 BPS loop still has one complex displacement and 
a superpartner. There should then be five complex tilts, as in the case of the 
1/3 BPS loop. The two doublets form multiplets $\{\bO^a,\bLambda^a\}$, 
$\{\Ob^a,\Lambdab^a\}$ and the singlet is now in a different multiplet 
$\{\bF,\bOb\}$. This last multiplet is not in the spectrum of the bosonic loop 
and this tilt generates motion along the $\bC\bP^1$, so we expect its 
normalisation $C_\bOb$, which starts as $C_\bDb/6$ at the 1/2 BPS 
point, to vanish as we approach the bosonic loop. Presumably there 
are still relations like  
$C_{\bO^a}+C_{\Ob_a}=C_\bDb/6$, as in the case of the 1/3 BPS loop 
studied here.

The fact that the singlet tilt is in the same multiplet with 
$\bF$ is consistent with the breaking the 1/2 BPS multiplet 
\eqref{1/2preserved1}, but cannot arise from the 1/3 BPS loop, 
where there are a pair $\bF$ and $\Fb$ in different multiplets 
without the singlet tilt. This is another indication that 
there is no 1/3 BPS loop in the 
same muduli space of 1/6 BPS loops based on $2\times2$ 
superconnections unrelated to 1/2 BPS loops.

Another family of 1/6 BPS loops that have previously not been studied 
are based on superconnections $\cL_1^+$ and $\cL_2^+$, where the 
latter, unlike $\cL_4^-$, is the direct $SU(4)$ rotation of $\cL_1^+$.
Unlike the 1/3 BPS loops, one can 
continuously rotate $W_1^+$ into $W_2^+$ while preserving 4 supercharges. 
These have the same bosonic symmetries as the generic 
1/6 BPS loops, so are a simpler setting to study this system and 
one can redo our calculation in Section~\ref{sec:DCM}, again relying 
on the 4-point functions that were calculated for the 1/2 BPS loop.

A better understanding of the space of line operators in the field 
theory could help in identifying the holographic 
duals, which is still an open question~\cite{Drukker:2008zx, 
Drukker:2019bev, Correa:2019rdk, Garay:2022szq}. To this we now add 
the puzzle of the holographic dual of the 1/3 BPS Wilson line. This is 
unlikely to be the 1/3 BPS solution of \cite{Drukker:2008jm}, but rather 
a superposition of two strings at different points on $\bC\bP^3$.

Other natural questions are the values of the normalisation constants 
(bremsstrahlung functions) for arbitrary 1/6 BPS loops, which are not 
bosonic. Likewise, one could push the analysis here to theories with 
$\cN=4$ supersymmetry~\cite{Gaiotto:2008sd, Imamura:2008dt}, 
and their equally rich spectrum of line 
operators~\cite{Assel:2015oxa, 
Ouyang:2015qma, Cooke:2015ila, Ouyang:2015iza, 
Ouyang:2015bmy, Mauri:2017whf, Dimofte:2019zzj, 
Drukker:2020dvr, Drukker:2022ywj, Drukker:2022bff}. We hope to 
address some of these questions, and many more that arise 
from this work, in the near future.

\section*{Acknowledgements}

We would like to acknowledge related collaboration with 
M. Probst, M. Tenser and D. Trancanelli. 
We are also grateful to G. Bliard, G. Papadopoulos, M. Preti, S. Salamon, 
V. Schomerus and P. Soresina for extremely useful discussions. 
ND is supported by STFC under the grants  ST/T000759/1 and ST/P000258/1 
and by the National Science Foundation under Grant No. NSF PHY-1748958. 
The work of ZK is supported by CSC grant No. 201906340174.
ND would like to thank \'Ecole Polytechnique F\'ed\'erale de Lausanne, 
the Kavli Institute for Theoretical Physics, UCSB, 
the DESY theory group and Humboldt University 
for their hospitality in the course of this work. ZK would like to thank 
T. Fiol, the University of Barcelona, the DESY theory group and 
Humboldt University for their hospitality.

\appendix

\section{Some 1/6 and 1/2 BPS Wilson lines}
\label{app:WLs}

We present here the BPS Wilson loops that are used in our analysis. All of them are 
straight lines along the $x_3$ axis denoted as $x$. The first is the bosonic Wilson 
loop~\cite{Drukker:2008zx,Chen:2008bp,Rey:2008bh} which is the ABJM analogue 
of the Gaiotto-Yin loop in $\cN=2$ theories~\cite{Gaiotto:2007qi}. 
It is 1/6 BPS, preserving the four supercharges 
${Q}^{12}_+$, $Q^{34}_-$, ${S}^{12}_+$, $S^{34}_-$. It is given as
\beq
\label{Wbos}
W_{\text{bos}}=\Tr \cP \exp\left(\int i\cA_{\mathrm{bos}} \,dx\right),
\eeq
where in the case of a loop in the first gauge group
\beq
\label{Mbos}
\cA_{\text{bos}}= A^{(1)}_x -\frac{2\pi i}{k} M\indices{^I_J} C_I \bar{C}^J\,,
\qquad M=\diag(-1,-1,1,1)\,,
\eeq
and similarly for the second group.

There is a large moduli space of Wilson loops preserving these 
supercharges~\cite{Drukker:2019bev, Drukker:2020dvr}. One first needs to 
elevate the bosonic Wilson loop to couple to both gauge groups as
\beq
\label{Wbos1}
W_{\text{bos}}=\Tr \cP \exp\left(\int i\cL_{\mathrm{bos}} \,dx \right),
\qquad
\cL_\text{bos}=\begin{pmatrix}
\cA_\text{bos}^{(1)}&0\\
0 & \cA_\text{bos}^{(2)}
\end{pmatrix},
\eeq
Then we can deform it as ($w_I$ and $\bar w^I$ are not necessarily 
complex conjugate)
\beq
\cL=\cL_\text{bos}-i({Q}^{12}_+ +Q^{34}_-)\cG+2\cG^2\,,
\qquad
\cG=\begin{pmatrix}
0&\bar w^I C_I\\
w_I\bar C^I&0
\end{pmatrix}.
\eeq
The explicit action of the supercharges on the fields is 
given in Appendix~\ref{app:SUSY}. The resulting loop is 1/6 BPS 
for arbitrary constant $w_1$, $w_2$, $\bar w^1$, $\bar w^2$ 
and the other vanishing or vice versa. Modding out by a $\bC^*$ 
action discussed in Section~\ref{sec:gauging}, 
the moduli space is two copies of the 
conifold~\cite{Drukker:2019bev, drukker:2020bps}.

In the case with $w_3=w_4=\bar w^3=\bar w^4=0$ we 
write those loops as in \eqref{Wbos1} with
\beq
\label{gen1/2}
\cL=\begin{pmatrix}
A_x^{(1)}+\alpha\bar\alpha M\indices{^I_J} C_I \bar{C}^J 
&-i\bar w^1\bar{\psi}_+^2+i\bar w^2\bar{\psi}_+^1\\
iw_1^+\psi_2^+-iw_2\psi_1^+
& A_x^{(2)} +\alpha\bar\alpha M\indices{^I_J} C_I \bar{C}^J
\end{pmatrix},
\quad
M\indices{^I_J}=(M_\text{bos})\indices{^I_J}+\frac{2}{\alpha\bar\alpha}\bar w^Iw_J\,,
\eeq
with $\alpha\bar\alpha=-2\pi i/k$. In the other case we have
\beq
\cL=\begin{pmatrix}
A_x^{(1)}+\frac{\alpha\bar\alpha}{2}{k}M\indices{^I_J} C_I \bar{C}^J 
&-i\bar w^3\bar{\psi}_-^4+i\bar w^4\bar{\psi}_-^3\\
-iw_3\psi_4^-+iw_4^-\psi_3^-
& A_x^{(2)} +\frac{\alpha\bar\alpha}{2}M\indices{^I_J} C_I \bar{C}^J
\end{pmatrix},
\quad
M\indices{^I_J}=(M_\text{bos})\indices{^I_J}+\frac{2}{\alpha\bar\alpha}\bar w^Iw_J\,.
\eeq

Within this space, the loops with $w_1\bar w^1+w_2\bar w^2=\alpha\bar\alpha$ are 1/2 BPS as 
are those with $w_3\bar w^3+w_4\bar w^4=-\alpha\bar\alpha$. 
The particular cases that are used in the body of the paper are:
\begin{itemize}
\item[\boldmath$W_1^+$:] 
Taking $w_2=\alpha$ and $\bar w^2=\bar\alpha$ 
satisfying $\bar{\alpha} \alpha=-2\pi i/k$, 
and all others vanishing we get
a loop with $SU(3)$ symmetry among indices 2, 3, 4
\beq
\label{L1}
\cL_1^+=\begin{pmatrix}
A_x^{(1)}+\alpha\bar\alpha (M_1)\indices{^I_J} C_I \bar{C}^J & i\bar\alpha\bar{\psi}_+^1\\
-i\alpha\psi_1^+ & A_x^{(2)} +\alpha\bar\alpha (M_1)\indices{^I_J} \bar{C}^JC_I 
\end{pmatrix},
\quad
M_1=\diag(-1,1,1,1)\,.
\eeq
Explicitly, $W_1^+$ preserves $Q^{12}_+$, $Q^{13}_+$, 
$Q^{14}_+$, $Q^{34}_-$, $Q^{24}_-$, $Q^{23}_-$ and the corresponding superconformal generators.

\item[\boldmath$W_4^-$:] 
The loop with $SU(3)$ symmetry among indices 1, 2, 3 has
\beq
\label{L4}
\cL_4^-=\begin{pmatrix}
    A_x^{(1)} +\alpha\bar\alpha (M_4)\indices{^I_J} C_I \bar{C}^J & i\bar\alpha \bar{\psi}_-^4\\
    -i\alpha  \psi_4^- & A_x^{(2)} +\alpha\bar\alpha  (M_4)\indices{^I_J} \bar{C}^J C_I
\end{pmatrix},
\quad
M_4=\diag(-1,-1,-1,1)\,.
\eeq
$W_4^-$ preserves $Q^{12}_+$, $Q^{13}_+$, $Q^{23}_+$, 
$Q^{34}_-$, $Q^{24}_-$, $Q^{14}_-$ and the corresponding $S$'s. 
It shares 8 supercharges with $W_1^+$.
\end{itemize}

\subsection{Supersymmetry transformations of the fields}
\label{app:SUSY}
Based on the conventions in \cite{Drukker:2022ywj} with 
some factors of 2 to be compatible with 
the algebra in Appendix~\ref{app:algebra} and with \cite{Bianchi:2020hsz}, 
the variations of the fields are
\bal
Q^{IJ}_\alpha C_K&=\delta^I_K\bar\psi^J_\alpha
-\delta^J_K\bar\psi^I_\alpha\,,
\\
Q^{IJ}_\alpha \bar C^K&=
-\epsilon^{IJKL}\epsilon_{\alpha\beta}\psi_L^\beta\,,
\\
Q^{IJ}_\alpha\psi_K^\beta&=
-2\delta^I_K\left(
i(\gamma^\mu)\indices{_\alpha^\beta}D_\mu\bar C^J
+2 \alpha\bar\alpha\delta_\alpha^\beta\bar C^{[J}C_L\bar C^{L]}\right)
\\&\quad
+2\delta^J_K\left(
i(\gamma^\mu)\indices{_\alpha^\beta}D_\mu\bar C^I
+ 2\alpha\bar\alpha\delta_\alpha^\beta\bar C^{[I}C_L\bar C^{L]}\right)
-8 \alpha\bar\alpha\delta_\alpha^\beta
\bar C^{[I}C_K\bar C^{J]}
\,,
\\
Q^{IJ}_\alpha \bar\psi^K_\beta&=
2\epsilon^{IJKL}\left(
i\epsilon_{\alpha\gamma}(\gamma^\mu)\indices{_\beta^\gamma}D_\mu C_L
+2\alpha\bar\alpha\epsilon_{\alpha\beta}C_{[L}\bar C^MC_{M]}\right)
\\&\quad
+4\alpha\bar\alpha\epsilon^{IJLM}\epsilon_{\alpha\beta}C_{[L}\bar C^KC_{M]}
\,,
\\
Q^{IJ}_\alpha A^{(1)}_\mu&=-\alpha\bar\alpha\epsilon^{IJKL}
\epsilon_{\alpha \gamma}(\gamma_\mu)\indices{_\beta^\gamma}C_K\psi_L^\beta
-2\alpha\bar\alpha(\gamma^\mu)\indices{_\alpha^\beta}
\bar\psi^{[I}_\beta\bar C^{J]}\,,
\\
Q^{IJ}_\alpha A^{(2)}_\mu&=\alpha\bar\alpha\epsilon^{IJKL}
\epsilon_{\alpha \gamma}(\gamma_\mu)\indices{_\beta^\gamma}\psi_K^\beta C_L
+2\alpha\bar\alpha(\gamma_\mu)\indices{_\alpha^\beta}
\bar C^{[I}\bar\psi^{J]}_\beta
\,.
\eal
The anti-symmetrisation symbol is normalised with a factor of $1/2$, 
the gamma matrices given by the Pauli 
matrices with $(\gamma^3)_+{}^+=1$ and $\epsilon^{+-}=\epsilon_{-+}=1$.

\section{Cusps, bremsstrahlungs and displacements}
\label{app:CBD}

In this appendix we review the necessary background on cusped Wilson loops, the small angle 
limit giving the bremsstrahlung functions and their relation to displacement and tilt operators.

\subsection{Cusped Wilson loops}
\label{app:cusps}
A cusped Wilson loop is comprised of two semi-infinite rays meeting at an angle $\phi$ such 
that $\phi=0$ is a straight line. We can parametrise the curve as
\beq
x^\mu(x)=\begin{cases}
(0,0,x)\,,&x<0\,,\\
(0,x\sin\phi,x\cos\phi)\,,&x>0\,.
\end{cases}
\eeq
Generically such loops suffer from logarithmic 
divergences~\cite{Polyakov:1980ca, Korchemsky:1987wg}, which means 
that the singular point obtains an anomalous dimension $\Gamma(\phi)$. For small 
angles this should be an even function, so to lowest order
\beq
\label{Bphi}
\Gamma(\phi)=-B_{\phi}\phi^2+\cO(\phi^4)\,,
\eeq
and $B^{\phi}$ is known as the bremsstrahlung function.

For loops coupling to scalar fields or fermions, we can also change those 
at the same point. With the structure of the 1/2 BPS loops, we can use 
the expressions in \eqref{gen1/2} and take
\beq
\left(w_1(x),w_2(x),\bar w^1(x),\bar w^2(x)\right)
=\begin{cases}
\alpha(0,-1, 0,-1)\,,&x<0\,,\\
\bar\alpha(\sin\frac{\theta}{2},-\cos\frac{\theta}{2},\sin\frac{\theta}{2},-\cos\frac{\theta}{2})\,,&x>0\,.
\end{cases}
\eeq
This would also lead to an anomalous dimension
\beq
\label{Btheta}
\Gamma(\theta)=B_{\theta}\theta^2+\cO(\theta^4)\,.
\eeq
More generally we have a function of both $\phi$ and $\theta$. 

One may wonder why the case of a straight line with a nonzero $\theta$ 
there is an anomaly, given that all the loops in \eqref{gen1/2} share 
four supercharges. The reason is that the supersymmetry variation of 
the superconnection $\cL$ does not vanish, but is a total derivative 
c.f. \eqref{total-deriv}. For different $\theta$, these total 
derivatives are different, so leave a boundary term at the location 
the change occures. In the special case of $\theta=\phi$ the two 
rays also share supercharges and the boundary terms cancel, so the 
combined system is BPS~\cite{Griguolo:2012iq}. In this case there 
should not be any anomaly and from the small angle expansions 
\eqref{Bphi} and \eqref{Btheta} we conclude that $B_\phi^{1/2}=B_\theta^{1/2}$.

For the bosonic loops \eqref{Wbos}, \eqref{Mbos} the $\phi$ cusp is as above 
and for the $\theta$ cusp we can take $M\indices{^I_J}$ along the second ray to be
\beq
M_\text{bos}^\theta=
\begin{pmatrix}-1&0&0&0\\
0&-\cos\theta&-\sin\theta&0\\
0&-\sin\theta&\cos\theta&0\\
0&0&0&1
\end{pmatrix}.
\eeq
In this case, $\phi=\theta$ is not a BPS configuration so the relation between $B^\text{bos}_\phi$ 
and $B^\text{bos}_\theta$ remained unclear until it was 
proven~\cite{Bianchi:2017afp, Bianchi:2017ujp, Bianchi:2018scb} that
\beq
B^\text{bos}_\phi=2B^\text{bos}_\theta\,.
\eeq

We are left with 
two independent functions $B^\text{bos}_\phi$ and $B^{1/2}_\phi$. 
The first is expressed as the derivative of 
the $n$-wound Wilson loop, which can be evaluated using 
localisation~\cite{Klemm:2012ii,Lewkowycz:2013laa}. The second 
can be related to the so-called latitude Wilson 
loop~\cite{Bianchi:2014laa, Bianchi:2017svd, Bianchi:2017ozk, 
Bianchi:2018bke}. They are also related via the 
framing anomaly factor that arises in 
calculating Wilson loops in Chern-Simons 
theory~\cite{Bianchi:2014laa,Bianchi:2016yzj}.

\subsection{Displacement and Twist}
\label{app:D&T}
The bremsstrahlung function of $\cN=4$ SYM in 4d was defined in~\cite{Correa:2012at}, 
where it was also related to the exact expectation value of the circular Wilson 
loop~\cite{erickson, gross, pestun} via the exact expectation value of other 
BPS Wilson loops~\cite{Drukker:2006ga, Drukker:2007yx, Drukker:2007dw}. 
The bremsstrahlung function is related to the two point functions of the 
displacement operator, and with enough supersymmetry, also of its 
superpartner, the tilt~\cite{Bianchi:2018zpb}.

In the context of ABJM theory, the relation between the bremsstrahlung 
function and the two point functions of displacement operators was presented 
in~\cite{Bianchi:2017ozk}. We do not repeat the derivation here, but 
the result is that the normalisation $C_\bD$ in \eqref{CD} is related to 
$B^\phi$ as%
\footnote{This is for the complex $\bD=\bD_1+i\bD_2$, so 
double the usual expression in~\cite{Correa:2012at}.}
\beq
\label{D-B}
C_\bD=24B_\phi\,.
\eeq
Such expression are valid for any BPS conformal loop, so the 1/2 BPS 
one, the bosonic loop and the 1/3 BPS loop as well.

A similar argument relates the two point function of the tilt $\bO$ to $B^\theta$. 
Specifically~\cite{Bianchi:2020hsz},
\beq
\label{O-B}
C_\bO=4B_\theta\,.
\eeq
In the case of the 1/2 BPS loop, this is consistent with $B_{1/2}^\phi=B_{1/2}^\theta$ 
and $C_\bD=6C_\bO$.

\section{Algebras and subalgebras}
\label{app:algebra}

We present here the superconformal algebra of ABJM and the 
subalgebras preserving various Wilson loops. We follow closely 
the notations in~\cite{Bianchi:2017ozk} so do not impose reality 
conditions. In~\cite{Bianchi:2017ozk} some factors of $i$ were 
introduced in describing the $\su(1,1|3)$ algebra. We refrain 
from doing that to avoid confusion and also do not introduce 
separate notations for the subalgebras associated to $W_1^+$ 
and $w_4^-$ or spell out their commutation relations, as they 
are all directly inherited from the original algebra. We 
could have imposed the reality condition on $\osp(6|4)$ that 
would be appropriate for a theory in $\bR^{2,1}$, but the 
benefit of that extra works seems marginal compared with 
consistency with~\cite{Bianchi:2017ozk, Bianchi:2020hsz}.

\subsection{Full $\osp(6|4)$ algebra}
\label{app:6|4}
The symmetry algebra of ABJM theory in flat space includes the conformal group 
$\sof(4,1)$, the R-symmetry $\su(4)$ and the supersymmetry generators forming 
the $\osp(6|4)$ superalgebra. The first is comprised of the 
Lorentz generators $M_{\mu\nu}$, translations $P_\mu$, special conformal transformations 
$K_\mu$ and dilation $D$. The R-symmetry generators can be written as $J\indices{_I^J}$ and 
we write them in a redundant notation allowing $I,J=1,\dots,4$, 
with the constraint $J\indices{_I^I}=0$. 
The supercharges are $Q^{IJ}_\alpha$ and $S^{IJ}_\alpha$ and satisfy the reality constraint 
$Q^{IJ}_\alpha=\frac12\epsilon^{IJKL}\bar Q_{KL\alpha}$ and 
$S^{IJ}_\alpha=\frac12\epsilon^{IJKL}\bar S_{KL\alpha}/2$.

The nonzero commutators in the conformal algebra are \cite{Zwiebel:2009vb,Papathanasiou:2009en}
\bal
{}[P_\mu,K_\nu]&=2\delta_{\mu\nu}D+2M_{\mu\nu}\,,
\quad&
[D,P_\mu]&=P_\mu\,,
\quad&
[D,K_\mu]&=-K_\mu\,,
\\
[M_{\mu\nu},M_{\rho\sigma}]&=\delta_{\mu[\sigma}M_{\rho]\nu}-\delta_{\nu[\sigma}M_{\rho]\mu}\,,
\quad&
[P_\mu,M_{\nu\rho}]&=\delta_{\mu[\nu}P_{\rho]}\,,
\quad&
[K_\mu,M_{\nu\rho}]&=\delta_{\mu[\nu}K_{\rho]}\,.
\eal
For the R-symmetry generators
\beq
[J\indices{_I^J},J\indices{_K^L}]=\delta_I^LJ\indices{_K^J}-\delta_K^JJ\indices{_I^L}\,.
\eeq
The anticommutators of the fermionic generators are
\bal
\label{QQ}
\{Q^{IJ}_\alpha,Q^{KL\beta}\}&=2\epsilon^{IJKL}(\gamma^\mu)\indices{_\alpha^\beta}P_\mu\,,
\qquad
\{S^{IJ}_\alpha,S^{KL\beta}\}=2\epsilon^{IJKL}(\gamma^\mu)\indices{_\alpha^\beta}K_\mu\,,
\\
\{Q^{IJ}_\alpha,S^{KL\beta}\}
&=\epsilon^{IJKL}(\gamma^{\mu\nu})\indices{_\alpha^\beta}M_{\mu\nu}
+2\delta_\alpha^\beta
\left(\epsilon^{IJKL} D-\epsilon^{NJKL}J\indices{_N^I}-\epsilon^{INKL}J\indices{_N^J}\right)
\,,
\eal
Finally the mixed commutators are
\begin{align}{}
\label{mixedcommutators}
[D,Q^{IJ}_\alpha]&=\frac12Q^{IJ}_\alpha\,,
\quad&
[D,S^{IJ}_\alpha]&=-\frac12S^{IJ}_\alpha\,,
\nonumber
\\
[M_{\mu\nu},Q^{IJ}_\alpha]&=-\frac12(\gamma_{\mu\nu})\indices{_\alpha^\beta}Q^{IJ}_\beta\,,
\quad&
[M_{\mu\nu},S^{IJ}_\alpha]&=-\frac12(\gamma_{\mu\nu})\indices{_\alpha^\beta}S^{IJ}_\beta\,,
\\\nonumber
[K_\mu,Q^{IJ}_\alpha]&=(\gamma_{\mu})\indices{_\alpha^\beta}S^{IJ}_\beta\,,
\quad&
[P_\mu,S^{IJ}_\alpha]&=(\gamma_{\mu})\indices{_\alpha^\beta}Q^{IJ}_\beta\,,
\\\nonumber
[J\indices{_I^J},Q^{KL}_\alpha]
&=\delta_I^KQ^{JL}_\alpha+\delta_I^LQ^{KJ}_\alpha-\frac12\delta_I^JQ^{KL}_\alpha\,,
\quad&
[J\indices{_I^J},S^{KL}_\alpha]
&=\delta_I^KS^{JL}_\alpha+\delta_I^LS^{KJ}_\alpha-\frac12\delta_I^JS^{KL}_\alpha\,.
\end{align}

\subsection{1/2 BPS $\su(1,1|3)$ subalgebras}
\label{app:2|3}

For the 1/2 BPS loop $W_1^+$, the preserved supercharges are $Q^{12}_+$, $Q^{13}_+$, $Q^{14}_+$, 
$Q^{23}_-$, $Q^{24}_-$, $Q^{34}_-$, and likewise $S^{12}_+$, etc.

Choosing $\gamma_3=\sigma_3$ and $(\sigma_3)\indices{_+^+}=1$, 
their anticommutators give the bosonic generators 
\beq
\label{algebra1}
P_3\,,\quad
K_3\,,\quad
M_{12}+2D\,,\quad
J\indices{_i^j}-\frac{1}{3}\delta_i^jJ\indices{_k^k}\,,\quad i,j,k\in\{2,3,4\}\,.
\eeq
Since $[P_3,K_3]=2D$, we get separately this generator and $M_{12}$. The full algebra can be easily 
read off from the commutators in Appendix~\ref{app:6|4} and can also be 
found in~\cite{Bianchi:2017ozk}.

Inside $\osp(6|4)$ there is an extra $\mathfrak{u}(1)$ symmetry $J\indices{_1^1}$ 
(or being pedantic about the tracelessness condition $J\indices{_1^1}-J\indices{_i^i}/3$) 
that commutes with this $\su(1,1|3)$. 
This generator acts nontrivially on the off-diagonal 
entries in $\cL_1^+$, the fermionic fields $\bar\psi^1_+$ and 
$\psi^+_1$. It's action on $\cL_1^+$ is the commutator with the 
supermatrix $T=\diag(I,-I)$ \eqref{T} (studied recently 
in~\cite{Gorini:2022jws}). $M_{12}$ has a similar action on the 
fermions, so the combination 
$M_{12}+J\indices{_1^1}/2$ acts trivially on the superconnection. 
Still each generator is a symmetry of the Wilson loop $W_1^+$, 
since their action either vanishes, or 
can be expressed as a total derivative of $\tau$, 
which integrates to zero \cite{Drukker:2009hy, Drukker:2022ywj}.

For the second 1/2 BPS loop, $W_4^-$, the preserved supercharges are $Q^{12}_+$, $Q^{13}_+$, $Q^{23}_+$, 
$Q^{14}_-$, $Q^{24}_-$, $Q^{34}_-$, and likewise $S^{12}_+$, etc.

Their algebra closes onto the bosonic generators
\beq
\label{algebra4}
P_3\,,\quad
K_3\,,\quad
D\,,\quad
M_{12}\,,\quad
J\indices{_{\hat\imath}^{\hat\jmath}}-\frac13\delta_{\hat\imath}^{\hat\jmath}J\indices{_{\hat k}^{\hat k}}\,,
\quad 
\hat\imath, \hat\jmath, \hat k\in\{1,2,3\}\,.
\eeq
And again, $M_{12}-J\indices{_4^4}/2$ generates an extra 
central $\mathfrak{u}(1)$ symmetry.

\subsection{1/3 BPS $\su(1,1|2)$ subalgebra}
\label{app:2|2}
The supercharges preserved by both $W_1^+$ and $W_4^-$ are $Q^{12}_+$, $Q^{13}_+$, 
$Q^{24}_-$, $Q^{34}_-$, and likewise $S^{12}_+$, etc.

Their algebra close onto $\su(1,1|2)$ and in particular the bosonic generators
\beq
P_3\,,\quad
K_3\,,\quad
D\,,\quad
M_{12}\,,\quad
J\indices{_2^3}\,,\quad
J\indices{_3^2}\,,\quad
J\indices{_2^2}-J\indices{_3^3}\,.
\eeq
Though not generated separately by the supercharges, the intersection 
of the two algebras \ref{algebra1} and \ref{algebra4} includes 
also $M_{1/2}+J\indices{_1^1}/2$ and $J\indices{_1^1}-J\indices{_4^4}$.

\subsection{Broken and unbroken generators}
\label{app:broken}

The table below lists all generators in $\osp(6|4)$ and whether they are broken by 
$W_1^+$, $W_4^-$ and/or $W_{1/3}$.
\begin{xltabular}{10cm}{@{}|l|c|c|c|@{}}
\hline
Generator&$W_1^+$&$W_4^-$& $W_{1/3}$\cr
\hline
$P_3$, $K_3$, $D$ &\cmark&\cmark&\cmark\cr
$P_1$, $P_2$, $K_1$, $K_2$ &\xmark&\xmark&\xmark\cr
$M_{12}$&\cmark&\cmark&\cmark\cr
$M_{13}$, $M_{23}$&\xmark&\xmark&\xmark\cr
$J\indices{_2^3}$, $J\indices{_3^2}$, $J\indices{_2^2}-J\indices{_3^3}$&\cmark&\cmark&\cmark\cr
$J\indices{_1^1}-\frac12J\indices{_2^2}-\frac12J\indices{_3^3}$, 
$J\indices{_4^4}-\frac12J\indices{_2^2}-\frac12J\indices{_3^3}$&\cmark&\cmark&\cmark\cr
$J\indices{_1^2}$, $J\indices{_1^3}$, $J\indices{_2^1}$, $J\indices{_3^1}$&\cmark&\xmark&\xmark\cr
$J\indices{_4^2}$, $J\indices{_4^3}$, $J\indices{_2^4}$, $J\indices{_3^4}$&\xmark&\cmark&\xmark\cr
$J\indices{_1^4}$, $J\indices{_4^1}$&\xmark&\xmark&\xmark\cr
$Q^{12}_+$, $Q^{13}_+$, $Q^{24}_-$, $Q^{34}_-$&\cmark&\cmark&\cmark\cr
$Q^{14}_+$, $Q^{23}_-$&\cmark&\xmark&\xmark\cr
$Q^{14}_-$, $Q^{23}_+$&\xmark&\cmark&\xmark\cr
$Q^{12}_-$, $Q^{13}_-$, $Q^{24}_+$, $Q^{34}_+$&\xmark&\xmark&\xmark\cr
$S^{12}_+$, $S^{13}_+$, $S^{24}_-$, $S^{34}_-$&\cmark&\cmark&\cmark\cr
$S^{14}_+$, $S^{23}_-$&\cmark&\xmark&\xmark\cr
$S^{14}_-$, $S^{23}_+$&\xmark&\cmark&\xmark\cr
$S^{12}_-$, $S^{13}_-$, $S^{24}_+$, $S^{34}_+$&\xmark&\xmark&\xmark\cr
\hline
\end{xltabular}

\section{Multiplet structure}
\label{app:commutators}
We list here the explicit action of the preserved supercharges on the 
tilt and displacement multiplets

\subsection{The tilt multiplets}
with $a=2,3$ and $\epsilon^{23}=-\epsilon_{23}=1$
\bal{}
\label{comm-bO}
\{Q^{1a}_+,\bF\}&=\bO^a\,,
&\quad
[Q^{1a}_+,\bO^b]&=\epsilon^{ab}\bLambda\,,
\\
%&&[S^{a4}_-,\bO^b]&=-2\epsilon^{ab}\bF\,,&\quad\{S^{a4}_-,\bLambda\}&=2\bO^a\,,\\
&&[Q^{a4}_-,\bO^b]&=2i\epsilon^{ab}\cD_x\bF\,,
&\{Q^{a4}_-,\bLambda\}&=-2i\cD_x\bO^a\,,
\eal

\bal{}
\label{comm-barbO}
\{Q^{a4}_-,\bar\bF\}&=\epsilon^{ab}\bar\bO_b\,,
&\quad
[Q^{a4}_-,\bar\bO_b]&=-\delta^a_b\bar\bLambda\,,
\\
%&&[S^{1a}_+,\bar\bO_b]&=2\delta^a_b\bar\bF\,,
%&\quad
%\{S^{1a}_+,\bar\bLambda\}&=-2\epsilon^{ab}\bar\bO_b\,,
%\\
&&[Q^{1a}_+,\bar\bO_b]&=2i\delta^a_b\cD_x\bar\bF\,,
&\{Q^{1a}_+,\bar\bLambda\}&=2i\epsilon^{ab}\cD_x\bar\bO_b\,,
\eal

\subsection{The tlit multiplets}
\bal{}
\label{tlit-algebra}
\{Q^{1a}_+,\Fb\}&=\epsilon^{ab}\Ob_b\,,
&\quad
[Q^{1a}_+,\Ob_b]&=-\delta^a_b\Lambdab\,,
\\
%&&[S^{a4}_-,\Ob_b]&=-2\delta^a_b\Fb\,,
%&\quad
%\{S^{a4}_-,\Lambdab\}&=-2\epsilon^{ab}\Ob_b\,,
%\\
&&[Q^{a4}_-,\Ob_b]&=-2i\delta^a_b\cD_x\Fb\,,
&\{Q^{a4}_-,\Lambdab\}&=-2i\epsilon^{ab}\cD_x\Ob_b\,,
\eal

\bal{}
\{Q^{a4}_-, \bar\Fb\}&= \bar\Ob^a\,,
&\quad
[Q^{a4}_-, \bar\Ob^b]&=-\epsilon^{ab} \bar\Lambdab\,,
\\
%&&
%[S^{1a}_+, \bar\Ob^b]&=2\epsilon^{ab} \bar\Fb\,,
%&\quad
%\{S^{1a}_+, \bar\Lambdab\}&=-2\bar\Ob^a\,,
%\\
&&[Q^{1a}_+, \bar\Ob^b]&=-2i\epsilon^{ab}\cD_x \bar\Fb\,,
&\{Q^{1a}_+, \bar\Lambdab\}&=-2i\cD_x \bar\Ob^a\,,
\eal

\subsection{Displacement multiplets}
\bal{}
[Q^{1a}_+,\bOb]&=-\epsilon^{ab}\bLambdab_b\,,
&\quad
\{Q^{1a}_+,\bLambdab_b\}&=-2\delta^a_b\bDb\,,
\\
%&&
%\{S^{a4}_-,\bLambdab_b\}&=2\delta^a_b\bOb\,,
%&\quad
%[S^{a4}_-,\bDb]&=-2\epsilon^{ab}\bLambdab_b\,,
%\\
&&\{Q^{a4}_-,\bLambdab_b\}&=2i\delta^a_b\cD_x\bOb\,,
&[Q^{a4}_-,\bDb]&=-i\epsilon^{ab}\cD_x\bLambdab_b\,,
\eal

\bal{}
[Q^{a4}_-,\bar\bOb]&=\bar\bLambdab^a\,,
&\quad
\{Q^{a4}_-,\bar\bLambdab^b\}&=-2\epsilon^{ab}\bar\bDb\,,
\\
%&&\{S^{1a}_+,\bar\bLambdab^b\}&=2\epsilon^{ab}\bar\bOb\,,&\quad[S^{1a}_+,\bar\bDb]&=-2\bar\bLambdab^a\,,\\
&&\{Q^{1a}_+,\bar\bLambdab^b\}&=-2i\epsilon^{ab}\cD_x\bar\bOb\,,
&[Q^{1a}_+,\bar\bDb]&=-i\cD_x\bar\bLambdab^a\,,
\eal

\section{Explicit expressions in terms of ABJM fields}
\label{app:fields}

\subsection{The $\sigma$ multiplet}
\label{app:sigma-mult}

The permutation operator $\sigma$ in the 1/3 BPS loop is given in 
\eqref{sigma}. Let us start with a more general $GL(2)$ matrix $g$ 
(or more precisly $g \in I_{2\times 2}\otimes GL(2)_{\mathbb{C}}$) 
acting by conjugation as
\beq
\begin{pmatrix}
\cL_1^+&0\\0&\cL_4^-
\end{pmatrix}
\to
g
\begin{pmatrix}
\cL_1^+&0\\0&\cL_4^-
\end{pmatrix}
g^{-1}
\eeq
To examine its variation under supersymmetry, we recall that~\cite{Drukker:2009hy}
\bal{}
\label{total-deriv}
[Q_+^{1a} ,i\cL_1^+]&=\cD_x^{\cL_1^+}\bar G^a\,,
\qquad&
[Q_+^{1a} ,i\cL_4^-]&=\cD_x^{\cL_4^-} G^a\,,
\\
[Q_-^{a4},i\cL_1^+]&=\cD_x^{\cL_1^+} G^a\,,
\qquad&
[Q_-^{a4},i\cL_4^-]&=\cD_x^{\cL_4^-}\bar G^a\,,
\eal
where
\beq
\label{G^a}
G^a=\begin{pmatrix}0& 2i\bar\alpha\epsilon^{ab} C_b\\0&0\end{pmatrix},
\qquad
\bar G^a=\begin{pmatrix}0&0\\ -2i\alpha\bar C^a&0\end{pmatrix},
\eeq
Integrating the total derivaties, we find the boundary terms
\beq
Q^{1a}_+W[g]=W\left[\begin{pmatrix}
    \bar G^a & 0\\
    0& G^a 
\end{pmatrix} g-g \begin{pmatrix}
    \bar G^a & 0\\
    0& G^a 
\end{pmatrix} \right]\,,
\eeq
as this is a local action, we can identify the action of the preserved 
supercharges on $g$ as
\bal
\label{Qsigma}
[Q^{1a}_+,g]&=\left[\begin{pmatrix}
    \bar G^a & 0\\
    0& G^a 
\end{pmatrix}, g\right],
\qquad
[Q^{a4}_-,g]&=\left[\begin{pmatrix}
    G^a & 0\\
    0& \bar{G}^a 
\end{pmatrix}, g\right].
\eal
A nicer action arises from the sum and difference of the supercharges
\bal{}
\label{Qsigma=0}
[Q^{1a}_++Q^{a4}_-,g]&=\left[\begin{pmatrix}
    \bar{G}^a +G^a & 0\\
    0 & \bar{G}^a +G^a
\end{pmatrix},g\right]=0
\\
[Q^{1a}_+ -Q^{a4}_-,g]&=\left[\begin{pmatrix}
    \bar{G}^a -G^a & 0\\
    0 & -(\bar{G}^a -G^a)
\end{pmatrix}, g\right]
= (\bar{G}^a -G^a)\otimes [\tau, g].
\eal
In the last expression we view $g$ as a $2\times2$ matrix and use the 
tensor symbol explicitly. We also use
\beq
\tau=\begin{pmatrix}
I&0\\0&-I
\end{pmatrix}.
\eeq
Clearly for a diagonal $g$ all variations cancel, so we can focus on 
off-diagonal $g$, which are linear combinations of $\sigma$ and 
$\tau\sigma$. We find then the descendents of $\sigma$ and $\tau\sigma$ as
\bal
\label{Sigma}
\bSigma^a&= \frac12[Q^{1a}_+ -Q^{a4}_-,\sigma]
=(\bar{G}^a -G^a)\otimes\tau\sigma\,,
\\
\Sigmab^a&= \frac12[Q^{1a}_+ -Q^{a4}_-,\tau\sigma]
=(\bar{G}^a -G^a)\otimes\sigma\,.
\eal

Looking at the second variation, first acting with the sum, 
then with the proper covariant derivative \eqref{Godd} we find
\bal
\label{Dsigma}
\frac12\{Q^{1a}_++Q^{a4}_-, \bSigma^b\}
&=\epsilon^{ab}\begin{pmatrix}
-2 \bar{\alpha} \alpha C_c \bar{C}^c
&i \bar{\alpha}(\bar{\psi}_-^4-\bar{\psi}_+^1)
\\-i \alpha(\psi_4^+-\psi_1^+) 
&-2 \bar{\alpha} \alpha \bar{C}^c C_c
\end{pmatrix}\otimes\tau\sigma
\\&
=\epsilon^{ab}
\begin{pmatrix}0&\cL_4^--\cL_1^+\\\cL_1^+-\cL_4^-&0\end{pmatrix}
=\epsilon^{ab}\cD_x\sigma\,.
\eal
Here $C_c \bar{C}^c=C_2\bar C^2+C_3\bar C^3$ and the result is
the covariant derivative of $\sigma$, in agreement with the 
algebra \eqref{QQ}. We find a similar result for $\Sigmab^a$.

Acting with the other combinations of supercharges we find the 
descendant
\bal
\label{o-def}
\epsilon^{ab}o
=\frac12\{Q^{1a}_+ - Q^{a4}_-, \bSigma^b\} 
&=\epsilon^{ab}\begin{pmatrix}
2\bar{\alpha} \alpha C_c \bar{C}^c
&0 \\0 & 
-2 \bar{\alpha} \alpha\bar{C}^c C_c 
\end{pmatrix}\otimes\sigma\\
&\quad-\epsilon^{ab}\begin{pmatrix}
0 &i \bar{\alpha}  (\bar{\psi}_+^1 +\bar{\psi}_-^4) \\
i \alpha(\bar{\psi}_+^1 +\bar{\psi}_-^4) &0
\end{pmatrix}\otimes\tau\sigma\,.
\eal
The expressions become a bit easier when starting with $\sigma^\pm=(\sigma\pm\tau\sigma)/2$
\bal
o^+&=\frac{1}{8}\{Q^{12}_+ - Q^{24}_-, \bSigma^3+\Sigmab^3\} 
=\begin{pmatrix}
2\bar{\alpha} \alpha C_c \bar{C}^c
&-i \bar{\alpha}  (\bar{\psi}_+^1 +\bar{\psi}_-^4)\\
-i \alpha(\bar{\psi}_+^1 +\bar{\psi}_-^4)& 
-2 \bar{\alpha} \alpha\bar{C}^c C_c 
\end{pmatrix}\otimes\sigma^+\,,
\\
o^-&=\frac{1}{8}\{Q^{12}_+ - Q^{24}_-, \bSigma^3-\Sigmab^3\} 
=\begin{pmatrix}
2\bar{\alpha} \alpha C_c \bar{C}^c
&i \bar{\alpha}  (\bar{\psi}_+^1 +\bar{\psi}_-^4)\\
i \alpha(\bar{\psi}_+^1 +\bar{\psi}_-^4)& 
-2 \bar{\alpha} \alpha\bar{C}^c C_c 
\end{pmatrix}\otimes\sigma^-\,.
\eal

\subsection{The tilt multiplets}
\label{app:tilt-fields}
We can act by the broken generators $J\indices{_1^a}$, 
$J\indices{_a^1}$, $J\indices{_a^4}$ and $J\indices{_4^a}$ on 
$\cL_1^+$ and $\cL_4^-$ to find the tilt operators
\bal
\label{O-fields}
\bO^a&=\begin{pmatrix}
-2 \bar{\alpha} \alpha C_1 \bar{C}^a & i\bar{\alpha} \bar{\psi}_+^a\\
0 & -2 \bar{\alpha} \alpha  \bar{C}^a C_1
\end{pmatrix},
&\quad 
\bar{\bO}_a&= \begin{pmatrix}
    2 \bar{\alpha} \alpha C_a \bar{C}^1 & 0\\
    i \alpha \psi_a^+ & 2 \bar{\alpha}\alpha \bar{C}^1 C_a
\end{pmatrix},
\\
\bar{\Ob}^a
&=\begin{pmatrix}
    2 \bar{\alpha} \alpha C_4 \bar{C}^a & i\bar{\alpha} \bar{\psi}_-^a \\
    0 & 2 \bar{\alpha} \alpha \bar{C}^a C_4
\end{pmatrix},
&\quad 
\Ob_a
&=\begin{pmatrix}
    -2 \bar{\alpha} \alpha C_a \bar{C}^4 & 0\\
    i \alpha \psi_a^- & -2 \bar{\alpha} \alpha \bar{C}^4 C_a
\end{pmatrix}.
\eal

By matching the fermionic parts of 
(see Appendix~\ref{app:commutators})
\bal
\{\tilde Q_+^{1a} ,\bF \}&=\bO^a\,,
\quad & 
\{\tilde Q_-^{a4} ,\bar{\bF}\} &=\epsilon^{ab} \bar{\bO}_b\,,
\\
\{\tilde Q_+^{1a},\Fb\} &=\epsilon^{ab} \Ob_b\,,
\quad & 
\{\tilde Q_-^{a4}, \bar{\Fb} \}&=\bar{\Ob}^a\,.
\eal
we get
\bal
\bF &=\begin{pmatrix}
    0 & i \bar{\alpha} C_1 \\
    0 & 0
\end{pmatrix},
&\quad 
\bar{\bF}&=\begin{pmatrix}
    0 & 0\\
    i \alpha \bar{C}^1 & 0
\end{pmatrix},
\\
\bar{\Fb}
&=\begin{pmatrix}
    0 & -i \bar{\alpha} C_4\\
    0 & 0
\end{pmatrix},
&\quad
\Fb&= \begin{pmatrix}
    0 & 0\\
    -i \alpha \bar{C}^4 & 0
\end{pmatrix}.
\eal
To make these expressions work, one needs to use the 
form of the variation on odd supermatrices in \eqref{Godd}.

The covariant supercharges acting on Grassmann odd matrices 
like $\bF$ and $\bar{\bF}$ inserted into $W_1^+$ as
(see the discussion in Appendix~\ref{app:subtlety})
\bal
\tilde Q_+^{1a} \bullet =Q_+^{1a} \bullet - \{\bar G^a, \bullet\}\,,
\quad 
\tilde Q_-^{a4} \bullet =Q_-^{a4} \bullet -\{ G^a , \bullet\}\,,
\eal
with $\bar G^a$ and $G^a$ in \eqref{G^a}. For the tlit operators 
inserted into $W_4^-$ we need to use the corresponding 
covariantisation with the roles of $\bar G^a$ and $G^a$ 
reversed.

We can then check that conversely (with a mixed anti-commutator 
for the even and odd entries in $\cL_1^+$)
\beq
[\tilde{Q}_-^{a4}, \bO^b] 
=2i \epsilon^{ab} (\partial_{x} \bF +i [\cL_1^+, \bF\})\,,
\eeq
and likewise should be the case for the other operators, 
in accordance with \eqref{comm-bO} and \eqref{comm-barbO}.

We can carry over the tlit operators $\Fb$, $\Ob_a$ and 
$\Lambdab$ to be insertions in $W_1^+$. We denote those 
operators as $\sigma\Fb\bar\sigma$, etc., but in terms of the 
field expressions, they have the same form as above. 
The difference is that when acting on them with a 
preserved charge, we need to use instead the appropriate 
covariantisation for $W_1^+$.

Since $Q^{1a}_++Q^{a4}_-$ annihilates $\sigma$ \eqref{Qsigma=0}, 
these operators have the same covariantisation with 
$G^a-\bar G^a$ inside both $W_1^+$ and $W_4^-$, so acting 
with them on $\sigma\bar{\Fb}\bar\sigma$ we find
\bal{}
\label{sigmaBbsigma}
[\tilde Q^{1a}_++\tilde Q^{a4}_-,\sigma\Fb\bar\sigma]
&=\epsilon^{ab}\sigma\Ob_b\bar\sigma\,,
\\
[\tilde Q^{1a}_++\tilde Q^{a4}_-,\sigma\bar\Fb\bar\sigma]
&=\sigma\bar\Ob^a\bar\sigma\,,
\eal
which is the appropriate covariantisation for operators in 
$W_4^-$.

Acting with $Q^{1a}_+$ and $Q^{a4}_-$ according to 
\eqref{comm-bO}, \eqref{comm-barbO} and the corresponding covariant derivatives \eqref{Geven}, 
we get the remainging operators in the tilt multiplets
\bal
\bLambda=-2 \bar{\alpha} \alpha \begin{pmatrix}
 \bar{\psi}_+^2 \bar{C}^3 - \bar{\psi}_+^3 \bar{C}^2 + C_1 \psi_4^- & -\frac{1}{\alpha} (D_1 -iD_2) C_4\\
 2 i \alpha  (\bar{C}^3 C_1 \bar{C}^2 - \bar{C}^2 C_1 \bar{C}^3) & \psi_4^- C_1 -\bar{C}^2 \bar{\psi}_+^3 + \bar{C}^3 \bar{\psi}_+^2
\end{pmatrix},\\
\bar{\bLambda} =-2\bar{\alpha} \alpha \begin{pmatrix}
    \bar{\psi}_-^4 \bar{C}^1 +C_2 \psi_3^+ -C_3 \psi_2^+ & 2 i \bar{\alpha} (C_3 \bar{C}^1 C_2 -C_2 \bar{C}^1 C_3)\\
    \frac{1}{\bar{\alpha}} (D_1+ iD_2) \bar{C}^4 &  \bar{C}^1 \bar{\psi}_-^4 + \psi_3^+ C_2 - \psi_2^+ C_3
\end{pmatrix}.
\eal
$D_1$ and $D_2$ are covariant derivatives (with the usual 
connections $A^{(1)}_\mu$ and $A^{(2)}_\mu$ in the transverse 
$\mu=1,2$ directions.

Likewise from $[Q_+^{1a},\Ob_b] =-\delta_b^a \Lambdab$ 
and $[Q_-^{a4}, \bar{\Ob}^b] =\epsilon^{ab} \bar{\Lambdab}$ we get
\bal
\Lambdab =-2\bar{\alpha} \alpha \begin{pmatrix}
\bar{\psi}_+^1 \bar{C}^4 +C_2 \psi_3^- -C_3 \psi_2^- & 2 i\bar{\alpha} (C_2 \bar{C}^4 C_3 - C_3 \bar{C}^4 C_2)\\
-\frac{1}{\bar{\alpha}} (D_1 -i D_2) \bar{C}^1 & \bar{C}^4 \bar{\psi}_+^1 +\psi_3^- C_2 - \psi_2^- C_3
\end{pmatrix},\\
\bar{\Lambdab} =-2\bar{\alpha} \alpha \begin{pmatrix}
    \bar{\psi}_-^2 \bar{C}^3 +C_4 \psi_1^+ -\bar{\psi}_-^3 \bar{C}^2 & \frac{1}{\alpha} (D_1 +iD_2) C_1\\
    2 i \alpha (\bar{C}^2 C_4 \bar{C}^3 -\bar{C}^3 C_4 \bar{C}^2) & \bar{C}^3 \bar{\psi}_-^2 + \psi_1^+ C_4 -\bar{C}^2 \bar{\psi}_-^3
\end{pmatrix}.
\eal

\subsection{The dispalcement multiplet}
\label{app:disp-fields}
It is easy to get the tilt operator in the displacement 
multiplet by replacing $a$ in \eqref{O-fields} with 4 and 1 we find
\bal
\bO^4&=\begin{pmatrix}
-2\bar{\alpha} \alpha C_1 \bar{C}^4 & i\bar{\alpha} \bar{\psi}_+^4\\
0 & -2\bar{\alpha} \alpha  \bar{C}^4 C_1
\end{pmatrix},
&\quad 
\bar{\bO}_4&= \begin{pmatrix}
    2\bar{\alpha} \alpha C_4 \bar{C}^1 & 0\\
    i \alpha \psi_4^+ & 2\bar{\alpha}\alpha \bar{C}^1 C_4
\end{pmatrix},
\\
\bar{\Ob}^1 &=\begin{pmatrix}
    2 \bar{\alpha} \alpha C_4 \bar{C}^1 & i\bar{\alpha} \bar{\psi}_-^1 \\
    0 & 2 \bar{\alpha} \alpha \bar{C}^1 C_4
\end{pmatrix},
&\quad 
\Ob_1 &=\begin{pmatrix}
    -2 \bar{\alpha} \alpha C_1 \bar{C}^4 & 0\\
    i \alpha \psi_1^- & -2 \bar{\alpha} \alpha \bar{C}^4 C_1
\end{pmatrix}.
\eal
$\bOb$ and $\bar{\bOb}$ are then expressed in terms of those 
as in \eqref{1/3disp}.

Then we find the explicit expressions for $\bLambda_a$, 
$\bar{\bLambda}^a$, $\Lambdab^a$ and $\bar\Lambdab_a$ as
\bal
\bLambda_a &=-2 \bar{\alpha} \alpha \begin{pmatrix}
    C_1 \psi_a^--\epsilon_{ab}(\bar{\psi}_+^b \bar{C}^4  -\bar{\psi}_+^4 \bar{C}^b) 
    & -\frac{1}{\alpha} (D_1 -iD_2) C_a\\
    -2i \alpha\epsilon_{ab} (\bar{C}^4 C_1 \bar{C}^b - \bar{C}^b C_1 \bar{C}^4) 
    & \psi_a^- C_1 -\epsilon_{ab}(\bar{C}^4 \psi_+^b -\bar{C}^b \bar{\psi}_+^4)
\end{pmatrix}\\
\bar{\bLambda}^a &=-2 \bar{\alpha} \alpha \begin{pmatrix}
    \bar{\psi}_-^a \bar{C}^1 -\epsilon^{ab}(C_4 \psi_b^+ + C_b \psi_4^+) & 
    2i \bar{\alpha}\epsilon^{ab} (C_4 \bar{C}^1 C_b -C_b \bar{C}^1 C_4)\\
    \frac{1}{\bar{\alpha}} (D_1 +i D_2) \bar{C}^a &
    \bar{C}^1 \bar{\psi}_-^a -\epsilon^{ab}(\psi_b^+ C_4 +\psi_4^+ C_b)
\end{pmatrix}
\eal
and
\bal
\Lambdab^a &=-2 \bar{\alpha} \alpha \begin{pmatrix}
    \bar{\psi}_+^a \bar{C}^4 -\epsilon^{ab}(C_1 \psi_b^- +C_b \psi_1^-) & 
    2i \bar{\alpha}\epsilon^{ab} (C_b \bar{C}^4 C_1 -C_1 \bar{C}^4 C_b)\\
    -\frac{1}{\bar{\alpha}} (D_1 -i D_2) \bar{C}^a & 
    \bar{C}^4 \bar{\psi}_+^a -\epsilon^{ab}(\psi_b^- C_1 +\psi_1^- C_b)
\end{pmatrix}\\
\bar{\Lambdab}_a &= -2 \bar{\alpha} \alpha \begin{pmatrix}
    C_4 \psi_a^+ -\epsilon_{ab}(\bar{\psi}_-^b \bar{C}^1 -\bar{\psi}_-^1 \bar{C}^b) & 
    \frac{1}{\alpha} (D_1 +i D_2) C_a\\
    -2i \alpha \epsilon_{ab}(\bar{C}^b C_4 \bar{C}^1 -\bar{C}^1 C_4 \bar{C}^b) & 
    \psi_a^+ C_4 -\epsilon_{ab}(\bar{C}^1 \bar{\psi}_-^b - \bar{C}^b \bar{\psi}_-^1)
\end{pmatrix}
\eal
The expressions for $\bD$ and $\Db$ can then be found by further action 
with the supercharges.

\subsection{Subtlety in covariant derivatives of supermatrices}
\label{app:subtlety}
For a Grassmann-even matrix $\cO$ inserted into a Wilson line
\bal
W[\cO(0)]=\Tr \cP \left[ \left(\exp \int_{-\infty}^0 i \cL(x) dx \right) \cO(0) \left(\exp \int_{0}^{\infty} i \cL(x) dx \right) \right],
\eal
as well as a Grassmann-even symmetry generator 
$\delta$ with $\delta (i\cL) =\cD_{x}^{\cL} \cG$, the 
variation of the Wilson line is
\bal
\delta W[\cO(0)]
&=W\left[ 
\left( \int_{-\infty}^0 \cD_{x}^{\cL} \cG(x^{\prime})\, dx^{\prime}\, \cO(0) 
+\delta \cO(0) 
+\cO(0) \int_0^{\infty} \cD_{x}^{\cL} \cG(x^{\prime})\, dx^{\prime} \right)
\right]
\\
&=W\left[\delta\cO +\cG \cO -\cO \cG)(0)\right].
\eal
So that we can define a covariant symmetry $\tilde{\delta}$ acting by
\bal
\label{covariantrelation}
\tilde{\delta} \cO =\delta\cO +\cG \cO -\cO \cG\,.
\eal

Turning to the case of Grassmann-odd operators, for example, the 
Grassmann-odd $Q$ and $G$
\bal
G=\begin{pmatrix}
    0 & g_{12}\\
    g_{21} & 0
\end{pmatrix},
\eal
with even $g_{ij}$. We can use a unit Grassmannian $\theta$ to repackage them 
into even objects $\delta= \theta Q$ and $\cG=\theta G$.

In the case where $\cO$ is an even supermatrix (like $\bO^a$ and $\bDb$)
\bal
\cO=\begin{pmatrix}
    B_{11} & F_{12}\\
    F_{21} & B_{22}
\end{pmatrix},
\eal
so the covariant action of $Q$ can be found from \eqref{covariantrelation} 
to be
\bal
\tilde{Q} \cO= \begin{pmatrix}
    Q B_{11} +g_{12} F_{21} +F_{12} g_{21} & Q F_{12} +g_{12} B_{22} -B_{11} g_{12}\\
    Q F_{21} +g_{21} B_{11} -B_{22} g_{21} & Q B_{22} +g_{21} F_{12} +F_{21} g_{12}
\end{pmatrix}.
\eal
In other words,
\bal
\label{Geven}
\tilde{Q} \cO= Q \cO  +\{G,\cO_{F}\} +[G, \cO_{B}]\,,
\eal
where $\cO_{B}$ and $\cO_{F}$ are the bosonic and fermionic parts of $\cO$.

The other case is for an odd supermatrix
\bal
\cO^{\prime} =\begin{pmatrix}
    F_{11} & B_{12}\\
    B_{21} & F_{22}
\end{pmatrix}.
\eal
We take an odd $\epsilon$ such that $\cO=\epsilon \cO^{\prime}$ 
is an even supermatrix. Then plugging this into 
\eqref{covariantrelation}, we get
\bal
\tilde{Q} \cO^{\prime} = \begin{pmatrix}
    Q F_{11} -(g_{12} B_{21} +B_{12} g_{21}) & Q B_{12} -(g_{12} F_{22} -F_{11} g_{12})\\
    Q B_{21} -(g_{21} F_{11} -F_{22} g_{21}) & Q F_{22} -(g_{21} B_{12} +B_{21} g_{12}).
\end{pmatrix}
\eal
In short
\bal
\label{Godd}
\tilde{Q} \cO^{\prime} = Q \cO^{\prime} - \{G, \cO^{\prime}_B\} -[G, \cO^{\prime}_F].
\eal

\section{The geometry of $SU(4)/S(U(2)\times U(1)\times U(1))$}
\label{app:coset}

As explained in Section~\ref{sec:DCM}, the defect conformal manifold is 
the quotient $SU(4)/S(U(2)\times U(1)\times U(1))$ and the integrated 4-point 
functions of the tilt operators are related to the curvature of this 
manifold. We follow~\cite{Mueller-Hoissen:1987cwl} 
(see also~\cite{Haupt:2017bnj}) to describe this quotient and evaluate 
the Riemann tensor.

We start by choosing explict generators of $SU(4)$ in terms of the  
$4\times4$ matrices, $\alpha_{ab}$ with entry 1 at location $ab$. The 
generators of $S(U(2)\times U(1)\times U(1))$ are the three diagonal ones and 
$\sqrt2\alpha_{12}$ and $\sqrt2\alpha_{21}$. Note that this is not a Hermitian 
basis, but we normalise them such multiplying by the hermitian conjugate and tracing 
gives 2. We denote them collectively as $h_A$ with $A=1,\dots,5$.

The remaining generators are
\bal
m_1&=\sqrt2\alpha_{13}\,,\quad&
m_2&=\sqrt2\alpha_{23}\,,\quad&
m_3&=\sqrt2\alpha_{41}\,,\quad&
m_4&=\sqrt2\alpha_{42}\,,\quad&
m_5&=\sqrt2\alpha_{43}\,,
\\
m_{\bar1}&=\sqrt2\alpha_{31}\,,\quad&
m_{\bar2}&=\sqrt2\alpha_{32}\,,\quad&
m_{\bar3}&=\sqrt2\alpha_{14}\,,\quad&
m_{\bar4}&=\sqrt2\alpha_{24}\,,\quad&
m_{\bar5}&=\sqrt2\alpha_{34}\,.
\eal
We denote them collectively as $m_i$. 
One can then define structure constants such that
\bal{}
[h_A,h_B]=f\indices{_A_B^C}h_C\,,
\qquad
[h_A,m_i]=f\indices{_A_i^j}m_j\,,
\qquad
[m_i,m_j]=f\indices{_i_j^A}h_A+f\indices{_i_j^k}m_k\,.
\eal

We do not wish to write explicit coordinates on the quo, but in any 
(local) representation in terms of group elemets $g$, 
the Maurer-Cartan form on the quotient can then be decomposed as
$g^{-1}dg=\ell^im_i+\Omega^Ah_A$. The metric on the quotient can then 
be written as
\beq
ds^2=g_{ij}\ell^i\ell^j\,,
\eeq
and this metric is $SU(4)$ invariant if $g_{AB}$ are constants and satisfy
\beq
f\indices{_A_i^k}g_{jk}+f\indices{_A_j^k}g_{ik}=0\,.
\eeq
In our case the possible solutions are
\beq
\label{coset-metric}
g_{1\bar1}=g_{\bar11}=g_{2\bar2}=g_{\bar22}=a\,,\qquad
g_{3\bar3}=g_{\bar33}=c\,,\qquad
g_{4\bar4}=g_{\bar44}=g_{5\bar5}=g_{\bar55}=b\,.
\eeq
In terms of the dCFT data \eqref{CO}, \eqref{CbO}, those are
\beq
a=C_{\bO_a}\,,\qquad
b=C_{\Ob^a}\,,\qquad
c=C_{\bOb}\,.
\eeq

The Levi-Civita connection is then given as
\beq
C\indices{_k^i_j}=\frac12\left(
g^{il}f\indices{_l_j^m}g_{km}
+g^{il}f\indices{_l_k^m}g_{jm}
+f\indices{_k_j^i}\right),
\eeq
And the Riemann tensor is
\beq
R\indices{^i_j_k_l}
=\left(
C\indices{_k^i_m}C\indices{_l^m_j}
-C\indices{_l^i_m}C\indices{_k^m_j}
-C\indices{_m^i_j}f\indices{_k_l^m}
-f\indices{_A_j^i}f\indices{_k_l^A}
\right).
\eeq
The full explanation of these expressions and their implementation 
for other quotients can be found in~\cite{Mueller-Hoissen:1987cwl}.

Lowering the first index and plugging in the metric and structure 
constants, we find that up to the usual symmetries of the Riemann tensor, 
the nonzero components of the form $R_{i\bar\jmath k\bar l}$ are
\bal
\label{explicit-R0}
&R_{1\bar11\bar1}=
R_{2\bar22\bar2}=
2a\,,
\qquad
R_{1\bar12\bar2}=
R_{1\bar22\bar1}=
a\,,
\\&
R_{1\bar13\bar3}=R_{2\bar23\bar3}=
-\frac{(a+b-c)^2-4ab}{4b}\,,
\\&
R_{1\bar14\bar4}=
R_{1\bar24\bar5}=
R_{2\bar15\bar4}=
R_{2\bar25\bar5}=
\frac{(a+b-c)^2-4ab}{4c}\,,
\\&
R_{1\bar44\bar1}=
R_{1\bar54\bar2}=
R_{2\bar45\bar1}=
R_{2\bar55\bar2}=
\frac{(a-b+c)(a-b-c)}{4c}\,,
\\&
R_{3\bar34\bar4}=
R_{3\bar35\bar5}=
-\frac{(a+b-c)^2-4ab}{4a}\,,
\\&
R_{1\bar33\bar1}=
R_{2\bar33\bar2}=
\frac{a+c-b}{2}\,,
\qquad
R_{3\bar44\bar3}=
R_{3\bar55\bar3}=
\frac{b+c-a}{2}\,,
\\&
R_{3\bar33\bar3}=2c\,,
\qquad
R_{4\bar44\bar4}=
R_{5\bar55\bar5}=
2b\,,
\qquad
R_{4\bar45\bar5}=
R_{4\bar55\bar4}=
b\,.
\eal
There are also nonvanishing $R_{ij\bar k\bar l}$ components
\begin{align}
\label{explicit-R2}
&
R_{1\bar31\bar3}=
R_{2\bar32\bar3}=
\frac{(a+b-c)(b+c-a)}{4b}\,,
\qquad
R_{3\bar43\bar4}=
R_{3\bar53\bar5}=
\frac{(a+b-c)(a+c-b)}{4a}\,.
\nonumber\\&
R_{14\bar1\bar4}=
R_{14\bar2\bar5}=
R_{25\bar1\bar4}=
R_{25\bar2\bar5}=
-\frac{a+b-c}{2}\,,
\end{align}
Such terms are incompatible with a K\"ahler structure and they all vanish 
for $\gamma=\alpha+\beta$, which is in fact the case for the 1/3 BPS 
loop \eqref{broken-ward}. With that condition also \eqref{explicit-R0} simplifies 
to
\bal
\label{explicit-R1}
&
\frac12R_{1\bar11\bar1}=
\frac12R_{2\bar22\bar2}=
%2a\,,
%\\&
R_{1\bar12\bar2}=
R_{1\bar22\bar1}=
R_{1\bar13\bar3}=
R_{2\bar23\bar3}=
R_{1\bar33\bar1}=
R_{2\bar33\bar2}=
a\,,
\\&
R_{1\bar14\bar4}=
R_{1\bar24\bar5}=
R_{2\bar15\bar4}=
R_{2\bar25\bar5}=
R_{1\bar44\bar1}=
R_{1\bar54\bar2}=
R_{2\bar45\bar1}=
R_{2\bar55\bar2}=
-\frac{ab}{a+b}\,,
\\&
R_{3\bar33\bar3}=2(a+b)\,,
\\&
R_{3\bar34\bar4}=
R_{3\bar35\bar5}=
R_{3\bar44\bar3}=
R_{3\bar55\bar3}=
R_{4\bar45\bar5}=
R_{4\bar55\bar4}=
\frac12R_{4\bar44\bar4}=
\frac12R_{5\bar55\bar5}=
b\,.
%\\&
%2b\,,
\eal

\bibliographystyle{utphys2}
\bibliography{refs}
\end{document}